 \documentclass[aps,prd,10pt,showpacs,amsmath,twocolumn,floatfix,amssymb,nofootinbib,longbibliography]{revtex4-2}
\usepackage{graphicx}
\usepackage[usenames]{color}
\usepackage{bm}
\usepackage{ifpdf}
\usepackage{floatrow}
\usepackage{makecell}
\usepackage{caption}

\usepackage{caption}
\usepackage{stackrel}
\usepackage{subcaption}
\usepackage{enumitem}
\usepackage[colorinlistoftodos]{todonotes}

\usepackage[normalem]{ulem}
\usepackage[dvipsnames]{xcolor}
\usepackage[utf8]{inputenc}
\usepackage{hyperref}
\hypersetup{
  pdfnewwindow=true,      
  colorlinks=true,        
  linkcolor=PineGreen,    
  citecolor=PineGreen,    
  filecolor=PineGreen,    
  urlcolor=PineGreen      
}
\newcommand{\be}{\begin{eqnarray}}
\newcommand{\ee}{\end{eqnarray}}

\begin{document}

\title{ Scattering phase shift in quantum mechanics on quantum computers}

\author{Peng~Guo}
\email{peng.guo@dsu.edu}

\affiliation{College of Arts and Sciences,  Dakota State University, Madison, SD 57042, USA}

\author{Paul~LeVan}
\email{paul.levan@dsu.edu}
 \affiliation{College of Arts and Sciences,  Dakota State University, Madison, SD 57042, USA}

\author{Frank~X.~Lee}
\email{fxlee@gwu.edu}
\affiliation{Department of Physics, The George Washington University, Washington, DC 20052, USA}

\author{Yong~Zhao}
\email{yong.zhao@anl.gov}
\affiliation{Physics Division, Argonne National Laboratory, Lemont, Illinois 60439, USA}

\date{\today}

\begin{abstract}
We investigate the feasibility of extracting infinite volume scattering phase shift on quantum computers in a simple one-dimensional quantum mechanical model, using the formalism established in Ref.~\cite{Guo:2023ecc}  that relates  the integrated correlation functions (ICF) for a trapped system to the infinite volume scattering phase shifts through a weighted integral. The system is first discretized in a finite box with periodic boundary conditions, and the formalism in real time is verified by employing a contact interaction potential with exact solutions. Quantum circuits are then designed and constructed to implement the formalism on current quantum computing architectures. To overcome the fast oscillatory behavior of the integrated correlation functions in real-time simulation, different methods of post-data analysis are proposed and discussed. Test results on IBM hardware show that good agreement can be achieved with two qubits, but complete failure ensues with three qubits due to two-qubit gate operation errors and thermal relaxation errors.     
\end{abstract}

\maketitle

\section{Introduction}\label{sec:intro}

 Scattering is an indispensable tool in our understanding of interactions in nature, from the original Rutherford experiment on the structure of the atom to modern experiments in nuclear and particle physics.  Theoretically, determination of scattering properties in hadronic systems from the first principles of quantum chromodynamics (QCD) remains fundamental but challenging. In most cases, numerical simulations based on Monte Carlo evaluation of the path integral are performed by placing the system in a finite volume with periodic boundary conditions, which leads to quantized energy spectrum in the system. The energy spectrum  is then  connected  to the infinite volume scattering phaseshifts through quantization conditions.
A number of finite-volume approaches have been proposed, including the well-known L\"uscher method \cite{Luscher:1990ux} which  has  proven successful in a wide range of applications~\cite{Aoki:2007rd,Beane:2007es, Detmold:2008fn,Feng:2010es,Lang:2011mn,Aoki:2011yj,Guo:2012hv,Guo:2016zos,Dudek:2016cru,Guo:2017ism,Guo:2018xbv,Guo:2019ogp,Mai:2019fba,Horz:2019rrn,Guo:2020kph,morningstar2025},
 the interaction potential method of HALQCD collaboration~ \cite{PhysRevLett.99.022001,PhysRevD.99.014514,ISHII2012437}, 
 and the integrated correlation function method (ICF)~\cite{Guo:2023ecc,Guo:2024zal,Guo:2024pvt,Guo:2025ngh,Guo:2025lmd,Guo:2025vgk}. The ICF method works directly with correlation functions, bypassing the energy spectrum determination in traditional methods.   Additional features include  the rapid  convergence at short Euclidean times  that makes it  potentially  a good candidate to overcome the S/N problem, and   free-by-construction from issues encountered at large volumes, such as increasingly dense energy spectrum and the extraction of low-lying states.  
 
 Computationally, all of the approaches are so demanding that lattice QCD simulations have been constantly pushing the limits of high-performance computing. 
 With the advent of quantum computing, it is natural for the field to explore its potential to speed up the simulations. In addition to larger lattices, near or at physical pion masses, and multi-baryon systems, quantum computing offers new prospects to  overcome intrinsic limitations facing classical simulations, such as no access to real-time dynamics,  and the sign problem in finite-density systems~\cite{PhysRevB.41.9301,deForcrand:2009zkb}.
Various ideas have been proposed in nuclear physics and lattice QCD in general~\cite{itou2025,Yamamoto_2024,Zhang_2021,Alexandru_2024}; and scattering in particular~\cite{davoudi2025,farrell2025,ingoldby2025,schuhmacher2025,Wang_2024,Li_2024,yusf2024,Sharma_2024,Brice_o_2021}.  
 Additionally,
there have been rapid developments recently on the subject of reaction processes in quantum simulations,  
including  (1)  combined  Variational Quantum Eigensolver    and   L\"uscher-  or BERW-like formula approach in Ref.~\cite{PhysRevC.109.064623,PhysRevC.110.054604};  (2) measuring phase shift by wave packet time delay \cite{PhysRevD.104.054507};   (3) reconstructing scattering amplitudes through the coupling of the particle with an ancillary spin-1/2 \cite{Mussardo2024}; (4) radiative processes \cite{Bedaque:2022ftd}; (5) inclusive process;  and other ideas, e.g. \cite{Gustafson:2020yfe,q36d-w649,Briceno:2020rar,Briceno:2023xcm}. 

The goal of the present work is to put the ICF formalism mentioned above to a realistic test  on the state-of-the-art quantum architectures, using the simplest formalism, namely, one-dimensional (1D) quantum mechanical model. 
We aim to explore the feasibility and practicality of extracting scattering phase shift in real time quantum simulation with ICF formalism, and the challenges and limitations facing current quantum computing hardware. We also propose and discuss different methods of post data analysis to overcome the fast oscillating behavior of integrated correlation functions in real time quantum simulation. Besides pedagogical value, the formalism and quantum circuits of 1D quantum mechanical model developed in this work 
set the baseline for more realistic scenarios with the ICF formalism on quantum computers, 
such as scalar field theory models~\cite{PhysRevA.99.052335,qr72-51v1}.

The paper is organized as follows. The quantum mechanical model setup,  a brief summary of the ICF formalism, and post-data analysis methods  are presented in Sec.~\ref{sec:scatt1DQM},  with some technical details given in appendices. Quantum circuits of the QM model are then detailed in Sec.~\ref{sec:QC}. Numerical tests on current quantum hardware   are discussed in Sec.~\ref{sec:numerics},   followed by a summary and outlook in Sec.~\ref{sec:summary}.

\section{Scattering in 1D Quantum Mechanical model}\label{sec:scatt1DQM}

Consider the Hamiltonian of 1D quantum mechanical model for a trapped particle interacting with a potential $V(x)$,
\begin{equation}
\hat{H} = \hat{H}_0  + V(x),
\end{equation}
where  
\begin{equation}
\hat{H}_0 =  - \frac{1}{2m}  \frac{d^2}{d x^2} + U(x),
\end{equation}
and  $U(x)$  denotes to the trap potential. We use a unit system with $\hbar=1$ and dimensionless distances. The common traps that are used in nuclear physics and lattice QCD communities include periodic box, harmonic oscillator trap, hard-wall boundary condition trap and uniform magnetic field trap, see e.g. Refs.~\cite{Guo:2021lhz,Guo:2021uig,Guo:2021qfu}. In the present work, we will primarily focus  on a  box of size $L$  with wave function satisfying periodic boundary condition
\begin{equation}
\psi(x+L) = \psi(x).
\end{equation}

\paragraph{Hamiltonian matrix in coordinate space: }  By discretizing $x$ into $x_\alpha = - \frac{L}{2} + a \alpha $, where $a = \frac{L}{N}$ is lattice spacing and $\alpha \in [0, 1, \cdots, N-1]$, and also replacing continuous derivative by 
\begin{equation}
\frac{d^2 \psi(x)}{d x^2}  \rightarrow  \frac{\psi(x+ a) - 2 \psi(x)  + \psi(x- a) }{a^2},
\end{equation}
the Hamiltonian matrix can be written as,
\begin{align}
\hat{H} & = - \frac{1}{2 m a^2}  \sum_{\alpha = 0}^{N-1}     \Bigl (\, | \alpha \rangle \langle \alpha +1 |  + | \alpha +1 \rangle \langle \alpha  |  \,\Bigr )   \nonumber \\
 & +  \sum_{\alpha = 0}^{N-1}  \left ( \frac{1}{ m a^2}  + V(x_\alpha ) \right )    | \alpha \rangle \langle \alpha  |      , \label{Hcoordinate}
\end{align}
where $| \alpha \rangle$ is a short-hand notation of $| x_\alpha \rangle$ basis. The periodic boundary condition requires $ | N \rangle = | 0 \rangle $, so that the particle is moving on a circle.
We will use a simple contact interaction potential $ V(x)=V_0 \delta(x) $ that yields the exact analytic solutions. Its discretized version is given by,
\begin{equation}
V(x_\alpha)  = \begin{cases}  \dfrac{V_0}{2 a }, & \alpha  \in \left[ \,\dfrac{N}{2}-1 ,  \dfrac{N}{2} \,\right] \\ 0, & \text{otherwise}  \end{cases}.  
\end{equation}
The continuum limit is approached by increasing the number of grid points and deceasing the lattice spacing simultaneously while keeping the box size fixed, 
\begin{equation}
\lim_{\substack{a\to 0\\ N\to \infty }} N a = L.
\end{equation}

\paragraph{Hamiltonian matrix in momentum basis: }  It is also convenient to work in the trapped basis without discretizing $x$. For a box of size $L$, the momentum basis that satisfies periodic boundary condition is,
\begin{equation}
\langle x | k \rangle = \frac{1}{\sqrt{L}} e^{i k x}, \ \ \ \ k = \frac{2\pi n}{L}, \ \ \ \ n \in \mathbb{Z}.
\end{equation}
The wave function can be expanded in terms of momentum basis as,
\begin{equation}
  | \psi \rangle = \sum_{k= \frac{2\pi n}{L}, n \in \mathbb{Z}}  \widetilde{\psi}  (k)  | k \rangle ,
\end{equation}
and the Hamiltonian matrix in momentum basis is given by a simple form,
\begin{equation}
\hat{H} = \sum_{(k,k')= \frac{2\pi n}{L}, n \in \mathbb{Z}}     \left [  \delta_{k, k'} \frac{k^2}{2m }  | k \rangle \langle k | + \frac{V_0}{L}  | k' \rangle \langle k |  \right ] . \label{Hmomentum}
\end{equation}

\subsection{Relating trapped dynamics to infinite volume scattering phase shift}

The ultimate goal is to compute scattering phase shift through quantum simulation. With a contact interaction potential in infinite volume, there exist analytic scattering solutions (see e.g. Refs.~\cite{Guo:2016fgl,Guo:2020kph,Guo:2020ikh,Guo:2020spn,Guo:2021uig,Guo:2022row}), as outlined in  Appendix~\ref{infvolumedynamics}.
The central result is that the infinite volume  two-particle elastic scattering phase shift, $\delta (\epsilon )$, is  related  to the integrated two-particle correlation function in a trap through a weighted integral,
\begin{equation}
C(t) - C_0 (t) \stackrel{\mbox{trap} \rightarrow \infty}{\longrightarrow} \frac{i t}{\pi} \int_0^{\infty} d \epsilon \delta(\epsilon) e^{- i \epsilon t},  \label{ICFmainEQ}
\end{equation}
 where $C(t)$ and $C_0 (t)$ refer to the integrated correlation functions for two non-relativistic particles interacting and non-interacting in the trap.  We will focus on non-relativistic dynamics in present work. For relativistic dynamics, extra kinematic factor must be taken into account, see Refs.~\cite{Guo:2024zal,Guo:2025lmd}.  
 The integrated correlation function can be computed by,
 \begin{equation}
 C(t) = Tr \left [ e^{- i \hat{H} t} \right ] = \sum_{n} e^{- i \epsilon_n  t},
 \end{equation}
where $\epsilon_n$'s are quantized eigen-energies of  the two-particle Hamiltonian in the trap. The non-interacting integrated correlation function, $C_0 (t)$, has the same form.   In addition to Eq.(\ref{ICFmainEQ}), the $n$-th moment of integrated correlation functions can also be computed
\begin{align} 
& Tr \left [ \left ( \hat{H} \right )^n  e^{- i \hat{H} t} -  \left ( \hat{H}_0 \right )^n  e^{- i \hat{H}_0  t}  \right ]  \nonumber \\
&  \stackrel{\mbox{trap} \rightarrow \infty}{\longrightarrow} i^n \frac{d^n    }{d t^n}  \left [  \frac{i t}{\pi} \int_0^{\infty} d \epsilon \delta(\epsilon) e^{- i \epsilon t} \right ].  \label{ICFmainEQmoment}
\end{align}

\begin{figure}
\includegraphics[width=0.95\textwidth]{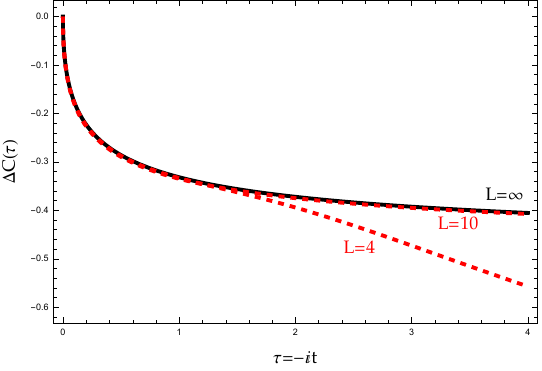}
\includegraphics[width=0.95\textwidth]{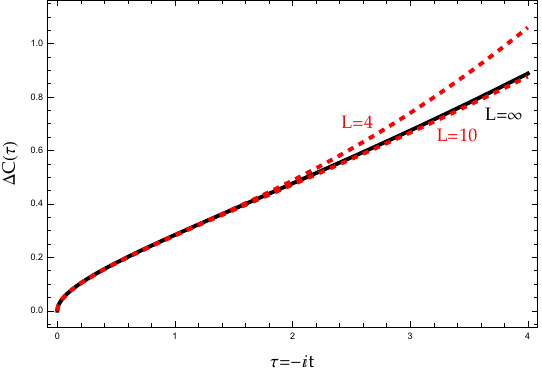}
 \caption{Convergence of Eq.(\ref{ICFmainEQanalytic}) 
 in Euclidean time with $L=4$ (dashed red), $10$ (dashed red) and $L = \infty$ (solid black). Both repulsive potential $V_0 =2$ (upper panel)  and attractive potential $V_0 = -0.5$ (lower panel) are demonstrated for parameters $m=1$ and $N=400$. The corresponding lattice spacing is $a = \frac{L}{N}=0.01,\, 0.025$. }  \label{citplot} 
 \end{figure}

\begin{figure}
\includegraphics[width=0.95\textwidth]{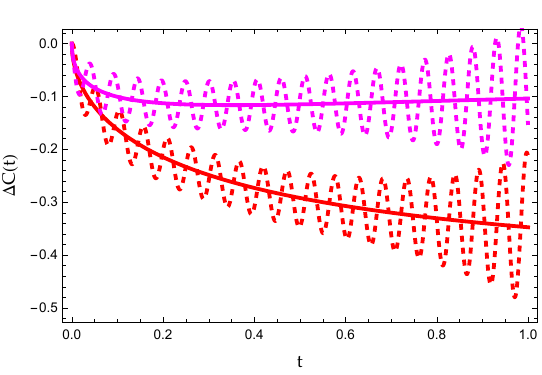}
 \caption{Demo plot of  real (red) and imaginary (purple) parts of $ \triangle C( t )  = C(t) - C_0 (t)$ in real time    vs. its infinite volume limit defined  in Eq.(\ref{ICFmainEQanalytic})  for a repulsive potential $V_0 =2$  with  $L=10$ (dashed) and $L = \infty$ (solid), where  $m=1$, $N=300$, and $a=0.033$.}
  \label{ctplot} 
 \end{figure}

With the contact interaction potential $V(x) = V_0 \delta(x)$, the analytic solution of infinite volume scattering phase shift is given by,
\begin{equation}
\delta(E) = \cot^{-1} \left ( - \frac{\sqrt{2 m E} }{m V_0} \right ).
\label{invPhase}
\end{equation}  
 Hence the analytic expression of right-hand side of Eq.(\ref{ICFmainEQ}) is available. With a repulsive potential ($V_0 >0$), we find,
 \begin{align}
 \frac{ i t}{\pi} \int_0^\infty \delta(\epsilon) e^{-  i\epsilon t} d \epsilon   =  \frac{1}{2} \mbox{erfc}  \left( m V_0  \sqrt{\frac{i t}{2m}}\right) e^{ (m V_0)^2 \frac{i t}{2m}} - \frac{1}{2} . 
\end{align}
 If the potential is attractive ($V_0 < 0$), bound state contribution must be added to the right-hand side of   Eq.(\ref{ICFmainEQ}),
 \begin{equation}
C(t) - C_0 (t) \stackrel{\mbox{trap} \rightarrow \infty}{\longrightarrow} \left ( e^{- i \epsilon_B t} -1 \right ) + \frac{i t}{\pi} \int_0^{\infty} d \epsilon \delta(\epsilon) e^{- i \epsilon t},  \label{ICFmainEQBound}
\end{equation}
 where the sole bound state energy for the contact interaction is,
 \begin{equation}
 \epsilon_B = - \frac{1}{2} m V_0^2. 
 \end{equation}
The corresponding expression for a attractive potential is given by,
 \begin{align}
& \frac{i t}{\pi} \int_0^\infty \delta(\epsilon) e^{- i  \epsilon t} d \epsilon \nonumber \\
&  = 
   -  \left ( \frac{1}{2} \mbox{erfc}  \left( |m V_0 | \sqrt{\frac{i t}{2m}}\right) e^{ (m V_0)^2 \frac{i t}{2m}} - \frac{1}{2}  \right ) . 
\end{align}
The combined result can be expressed in terms of the complementary error function, $\mbox{erfc} (x) = 1- \mbox{erf}(x)$,
 \begin{equation}
C(t) - C_0 (t) \stackrel{\mbox{trap} \rightarrow \infty}{\longrightarrow}  \frac{1}{2} \mbox{erfc}  \left( m V_0  \sqrt{\frac{i t}{2m}}\right) e^{ (m V_0)^2 \frac{i t}{2m}} - \frac{1}{2}  ,  \label{ICFmainEQanalytic}
\end{equation}
which is valid for both repulsive and attractive potentials. The demo plots of $\triangle C(\tau) =C(\tau ) -C_0 (\tau)$ in Euclidean time $\tau = - i t$ vs.  its infinite volume limit defined  in Eq.(\ref{ICFmainEQanalytic})  for both repulsive and attractive potentials are shown in Fig.~\ref{citplot}. 
The demo plot of $\triangle C(t) =C(t ) -C_0 (t)$ in real time vs.  its infinite volume limit defined  in Eq.(\ref{ICFmainEQanalytic}) for a repulsive potential is shown in  Fig.~\ref{ctplot}.

\subsection{Post data analysis}\label{sec:postdataanalysis}

In Euclidean (or imaginary) time evolution, the  left-hand side of Eq.(\ref{ICFmainEQ}) approaches to its infinite volume limit on the right-hand side of Eq.(\ref{ICFmainEQ}) rapidly when the size of trap is increased, as shown in Fig.~\ref{citplot}.
On the other hand, for real time evolution,  the integrated correlation functions $C(t)$ and $C_0 (t)$ in the trap exhibit fast oscillating behavior around its infinite volume limit due to quantized eigen-solutions of the trap dynamics, see Fig.~\ref{ctplot}. This presents challenges on quantum computers.  Post-data analysis must be deployed in real time simulation of the integrated correlation functions in order to extract scattering phase shift from fast oscillating data of $C(t) - C_0 (t)$. 

One idea is to smooth out   the fast oscillatory behavior by averaging out over a short period of time that is larger than oscillation period, as was suggested in Ref.~\cite{Guo:2025vgk} (also see Ref.\cite{Burbano:2025pef}). 
In the present work, we explore other possibilities of post-data analysis.   Using the integral transform $ - i \int_0^\infty d t e^{i E t} \left  [\cdots \right ] $, Eq.(\ref{ICFmainEQ}) can be brought into a compact form amenable to numerical verification,
 \begin{equation}
 Tr \left [ \frac{1}{E- \hat{H}}  -  \frac{1}{E- \hat{H}_0}  \right ]    \stackrel{\mbox{trap} \rightarrow \infty}{\longrightarrow}     -  \frac{d}{d E}  \ln T(E)  , \label{dlogTdE}
\end{equation}
where $T(E)$ in 1D quantum mechanics is identified as the transmission amplitude and can be parameterized through scattering phase shift by,
\begin{equation}
T(E)   = \frac{e^{2 i \delta (E)} + 1}{2}  = \cos \delta(E) e^{ i\delta (E)}. \label{Tphaseshiftparam}
\end{equation}
Eq.(\ref{dlogTdE}) is known as Friedel formula (or  Krein's theorem) in formal scattering theory. Its proof is rather involved and an outline can be found in Appendix \ref{Fridelformula}.
With the help of Eq.(\ref{Tphaseshiftparam}),   we find the relation,
 \begin{equation}
 Tr \left [ \frac{1}{E- \hat{H}}  -  \frac{1}{E- \hat{H}_0}  \right ]    \stackrel{\mbox{trap} \rightarrow \infty}{\longrightarrow}       \left [ \tan  \delta (E)   - i  \right ]  \frac{d \delta (E)}{d E} . 
\end{equation}
  The phase shift can be computed through,
 \begin{equation}
   \delta (E)  = \lim_{\mbox{trap} \rightarrow \infty} \phi (E)     , \label{phaseshift}
\end{equation}
where the phase angle $\phi(E)$ is defined for the trapped system by,
 \begin{equation}
 \phi (E)  =  \cot^{-1} \left ( - \frac{Im \left [ \triangle \widetilde{C} (E)  \right ]}{Re \left [ \triangle \widetilde{C} (E)  \right ]} \right )     . \label{phaseintrap}
\end{equation}
 The $\triangle \widetilde{C} (E) $ is the difference of integral transformed integrated correlation functions, 
 \begin{equation}
 \triangle \widetilde{C} (E) =   \widetilde{C} (E) -  \widetilde{C}_0 (E)   ,
 \end{equation}
 where
 \begin{equation}
   \widetilde{C} (E)  =  - i \int_0^\infty d t e^{i E t}  C(t) = Tr \left [ \frac{1}{E- \hat{H}}    \right ]  = \sum_{n} \frac{1}{E- \epsilon_n},
 \end{equation}
and $ \widetilde{C}_0 (E)  $ is defined  similarly.

  \begin{figure}
\includegraphics[width=0.95\textwidth]{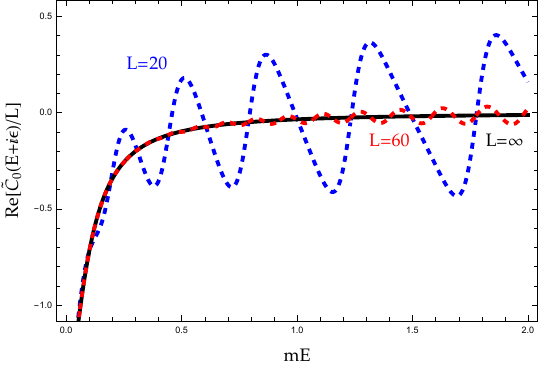}
\includegraphics[width=0.95\textwidth]{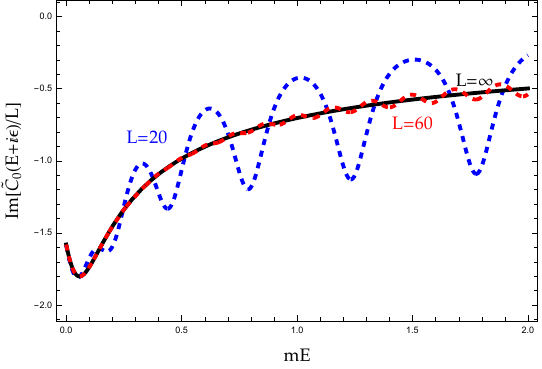}
 \caption{Convergence in the $E+i \varepsilon$ prescription for real (upper panel) and imaginary (lower panel) parts of $ \frac{1}{L} \widetilde{C}_0(E + i \varepsilon) $ defined in Eq.(\ref{C0Lexample}) vs. its infinite volume limit defined in Eq.(\ref{C0infexample}) for $L=20$ (dashed blue), $60$ (dashed red) and $L = \infty$ (solid black), where $ \varepsilon = 0.1$ and $m=1$.
 }  \label{Gfvplot} 
 \end{figure}

Although Eq.(\ref{phaseshift}) and Eq.(\ref{phaseintrap}) are formally correct at infinite volume limit,  their application to the  extraction of phase shift from a trap of finite size requires extra caution. In a finite trap, the discrete eigen-energies   create isolated poles sitting along the real axis in the complex energy plane in $\widetilde{C} (E) $ and $\widetilde{C}_0(E)$ functions. Hence with real $E$ values,  both $\widetilde{C} (E) $ and $\widetilde{C}_0(E)$ are real oscillating functions that diverge at pole positions. For example, in a periodic box with size of $L$, the discrete eigen-energies for free particle are $\epsilon_n^{(0)}  = \frac{1}{2m} ( \frac{2\pi n}{L} )^2$ where $n \in \mathbb{Z}$, and 
\begin{equation}
 \frac{1}{L} \widetilde{C}_0(E) = \frac{1}{L} \sum_{n \in \mathbb{Z}} \frac{1}{E- \epsilon_n^{(0)}}  = m  \frac{\cot \left ( \frac{\sqrt{2 m E} L}{2} \right )}{\sqrt{2 m E}} . \label{C0Lexample}
\end{equation}
At the infinite volume limit,   $\frac{1}{L } \sum_n $  is replaced by $\int_{-\infty}^{\infty} \frac{d p}{2\pi}$  and the poles dissolve into branch cuts sitting along the real axis, leading to,
\begin{equation}
 \frac{1}{L} \widetilde{C}_0(E)  \stackrel{\mbox{trap} \rightarrow \infty}{  \longrightarrow } \int_{-\infty}^{\infty}   \frac{d p}{2\pi}  \frac{1}{E- \frac{p^2}{2m} }  = -  m \frac{i}{ \sqrt{2 m E}}  \label{C0infexample},
\end{equation}
which is a smooth pure imaginary function, also see detailed discussion in Ref.~\cite{Guo:2020ikh}.

To regulate the divergence of poles in trapped functions $\widetilde{C} (E) $ and $\widetilde{C}_0(E)$ and to smooth out the sharp oscillating behavior, we consider two remedies.

\begin{itemize}
\item $E+i \varepsilon$ prescription:   a finite imaginary part $ i \varepsilon$ is added to energy. To match  $\widetilde{C} (E) $ or $\widetilde{C}_0(E)$ with their infinite volume limits, $\varepsilon$ is chosen to satisfy $\sqrt{ \varepsilon  } \gg 1/L$, see Fig.~\ref{Gfvplot} as a example.

\item $ L \rightarrow i L  $ rotation:   Another useful technique in finite volume to smooth out   the oscillating behavior  of correlation function is to   rotate finite volume $L$ to imaginary axis $i L$, see detailed discussion in Ref.~\cite{Guo:2020ikh}.   After rotation, 
\begin{align}
 &\frac{1}{L} \widetilde{C}_0(E)  \stackrel{L \rightarrow i L}{\longrightarrow} \frac{1}{ i L} \sum_{n \in \mathbb{Z}} \frac{1}{E+ \epsilon_n^{(0)}}  \nonumber \\
& = - m  \frac{i }{\sqrt{2 m E}}  \label{C0iLexample} \coth \left ( \frac{\sqrt{2 m E} L}{2} \right ),
\end{align}
where 
\begin{equation}
\coth \left ( \frac{\sqrt{2 m E} L}{2} \right )  \stackrel{L \rightarrow \infty}{\longrightarrow} 1.
\end{equation}
Hence the $i L$-rotated  $ \widetilde{C}_0(E)/L$ approaches its infinite volume limit rapidly. 
\end{itemize}

We remark that in 1D quantum mechanics, it seems appealing to compute   $ \widetilde{C} (E)  = Tr \left [ \frac{1}{E- \hat{H}}    \right ]$ directly  by matrix inversion in order to extract scattering phase shift using relations given by Eq.(\ref{phaseshift}) and Eq.(\ref{phaseintrap}).  Existing matrix inversion quantum algorithms include the Harrow–Hassidim–Lloyd (HHL) algorithm \cite{PhysRevLett.103.150502} and  the  Quantum Singular Value Transform (QSVT) algorithm \cite{AndrewM:2015yvx,PRXQuantum.2.040203}, both of which can be applied to non-Hermitian matrices.  
However, it is not clear that working with the Hamiltonian matrix is the most effective approach when it comes to quantum computing of quantum field theory models. It is a topic worth exploring. Here we will focus on the time evolution of correlation functions since they are are widely used physical quantities across a variety of subfields in physical sciences, including nuclear/particle physics and condensed matter physics.

  \begin{figure}[!htbp]
\includegraphics[width=0.9\textwidth]{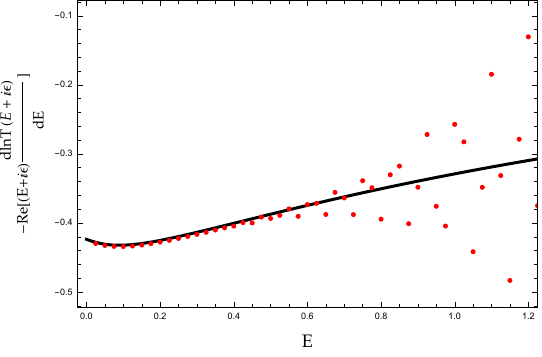}
\includegraphics[width=0.9\textwidth]{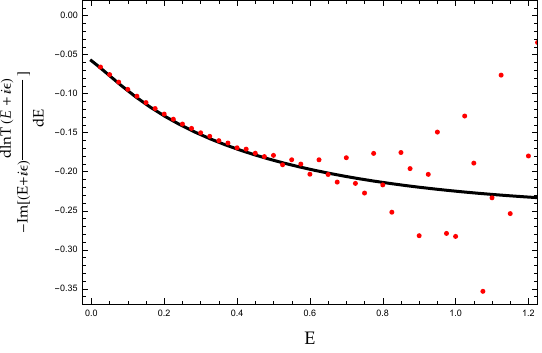}
\includegraphics[width=0.9\textwidth]{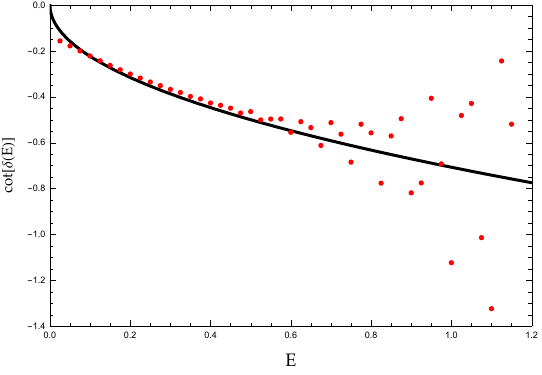}
 \caption{Verification of Eq.\eqref{dlogTdE} in the $E + i \varepsilon$ prescription: real (top) and imaginary (middle) parts of  $ (E+ i\varepsilon) \triangle \widetilde{C}(E+ i \varepsilon) $ (red dots) for a periodic trap vs. its infinite volume limit  $- (E + i \varepsilon) \frac{d}{d E} \ln T(E+ i \varepsilon)$ (solid black)  with $L=100$ and $ \varepsilon = 0.1$.  
 Bottom: comparison of  $\cot   \phi(E + i \varepsilon) $ (red dots) in Eq.(\ref{phaseshift}) and Eq.(\ref{phaseintrap}) vs. infinite volume limit phase shift $\cot  \delta(E) $ (solid black) in Eq.\eqref{invPhase}. The rest of the parameters are: $V_0 =2$, $m=1$, $N=4000$ and $a \sim 0. 025$.}  \label{CEplot} \vspace*{-0.1in}
 \end{figure}

   \begin{figure}
\includegraphics[width=0.95\textwidth]{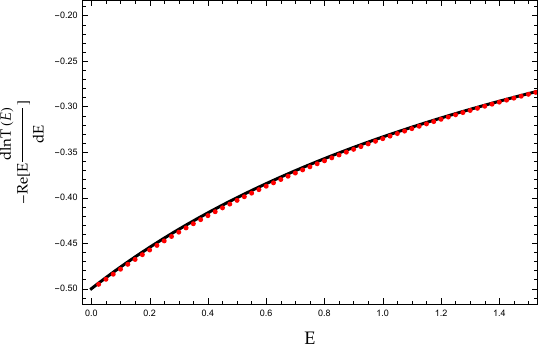}
\includegraphics[width=0.95\textwidth]{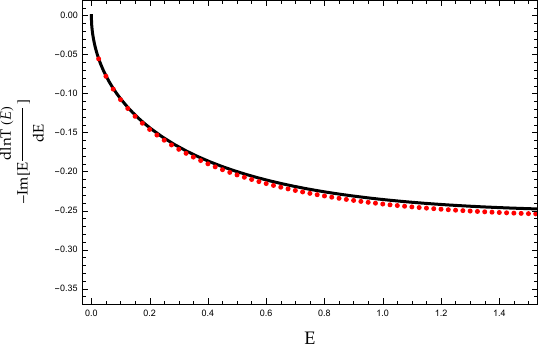}
\includegraphics[width=0.89\textwidth]{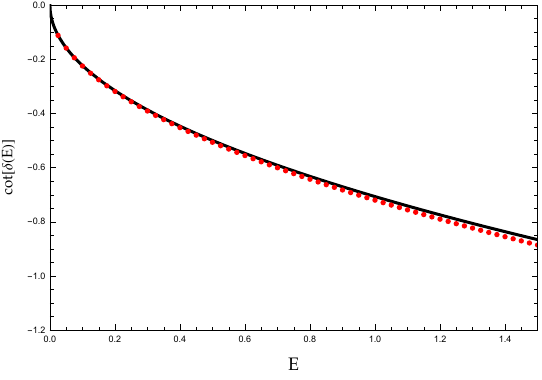}
 \caption{Similar to Fig.~\ref{CEplot}, but in the $ L \rightarrow i L  $ prescription  with $L=100 i$.}  \label{CEiLplot} 
 \end{figure}

 Having investigated the pole structure of the integral- transformed ICFs $\widetilde{C}(E)$ and $\widetilde{C}_0(E)$,
 we are now ready to verify the relations in Eq.(\ref{dlogTdE}),  Eq.(\ref{phaseshift}) and Eq.(\ref{phaseintrap}) using analytic solutions of the contact interaction potential model in both prescriptions.

In the $E + i \varepsilon$ prescription,   we choose to multiply Eq.\eqref{dlogTdE} by $E$ and plot $  E \triangle \widetilde{C}(E )  = Tr \left [ \frac{\hat{H}}{E- \hat{H}} -  \frac{\hat{H}_0 }{E- \hat{H}_0} \right ]$ in a periodic box for $L=100$ and $a \sim 0.025$ vs.  its infinite volume limit  $- E \frac{d}{d E} \ln T(E )$  with $\varepsilon=0.1$ in the top two panels of Fig.~\ref{CEplot}. In the bottom panel of the same plot, we show the comparison of  $\cot \phi(E + i \varepsilon)$ phase in the trap defined in Eq.(\ref{phaseintrap})  vs. infinite volume scattering phase shift $\cot \delta (E)$ in Eq.\eqref{invPhase}.
The agreement is fairly good up to around $E=0.5$, then starts to deteriorate significantly, despite the use of a fairly large box ($L=N a$). 

In the $ L \rightarrow i L  $ rotation method, 
the rotated Hamiltonian matrix becomes non-Hermitian,
\begin{align}
\hat{H} &  \stackrel{ L \rightarrow iL }{\longrightarrow }  \frac{1}{2 m a^2}  \sum_{\alpha = 0}^{N-1}     \Bigl (\, | \alpha \rangle \langle \alpha +1 |  + | \alpha +1 \rangle \langle \alpha  |  \,\Bigr )   \nonumber \\
 & -  \sum_{\alpha = 0}^{N-1}  \left (  \frac{1}{ m a^2}  + i  V(x_\alpha ) \right )    | \alpha \rangle \langle \alpha  |      .
\end{align}
This is to be compared with the Hermitian version in Eq.\eqref{Hcoordinate}.
The numerical verification in the $ L \rightarrow i L  $ prescription is given in Fig.~\ref{CEiLplot}. 
Unlike the $ E + i \varepsilon$ prescription in Fig.~\ref{CEplot}, excellent agreement is observed over the entire energy range plotted.

\section{Quantum circuits}\label{sec:QC}
Having verified how the ICF formalism works in 1D quantum mechanics, we now set out to implement it on quantum computers.
The Hamiltonian matrices in Eq.(\ref{Hcoordinate}) and Eq.(\ref{Hmomentum}) can be mapped into quantum circuits rather straightforwardly. A $N \times N$ matrix can be mapped on $\Gamma = \log_2 N$ quantum registers. In the following we present the complete quantum circuits for the Hamiltonian matrices in Eq.(\ref{Hcoordinate}) and Eq.(\ref{Hmomentum}), some of which have been discussed in Refs.~\cite{Guo:2025vgk,Guo:2025xpd}.

\begin{figure}[h]
\frame{\includegraphics[width=.59\textwidth]{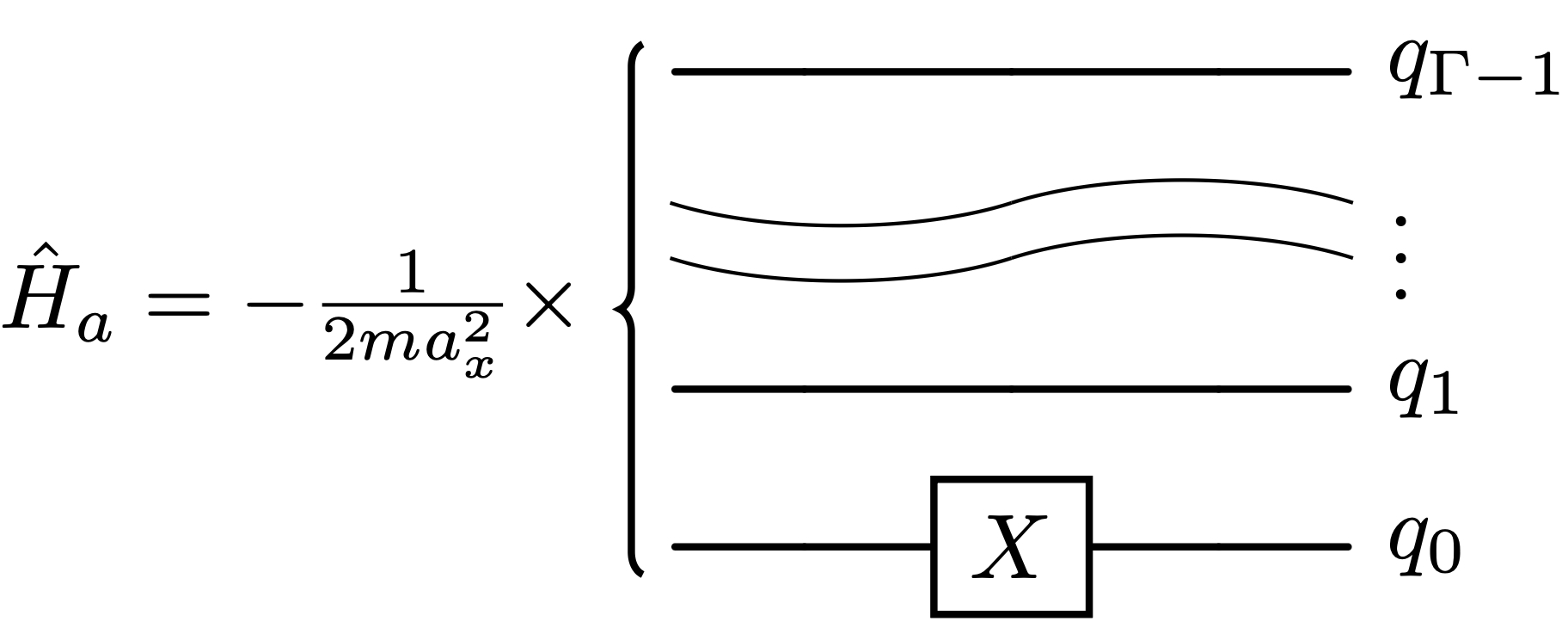}}
\caption{Quantum circuit of $\hat{H}_a$ in Eq.\eqref{HaHbHvcoordinate}.}
\label{FIGHa}
\end{figure}

\begin{figure}[h]
\frame{\includegraphics[width=.99\textwidth]{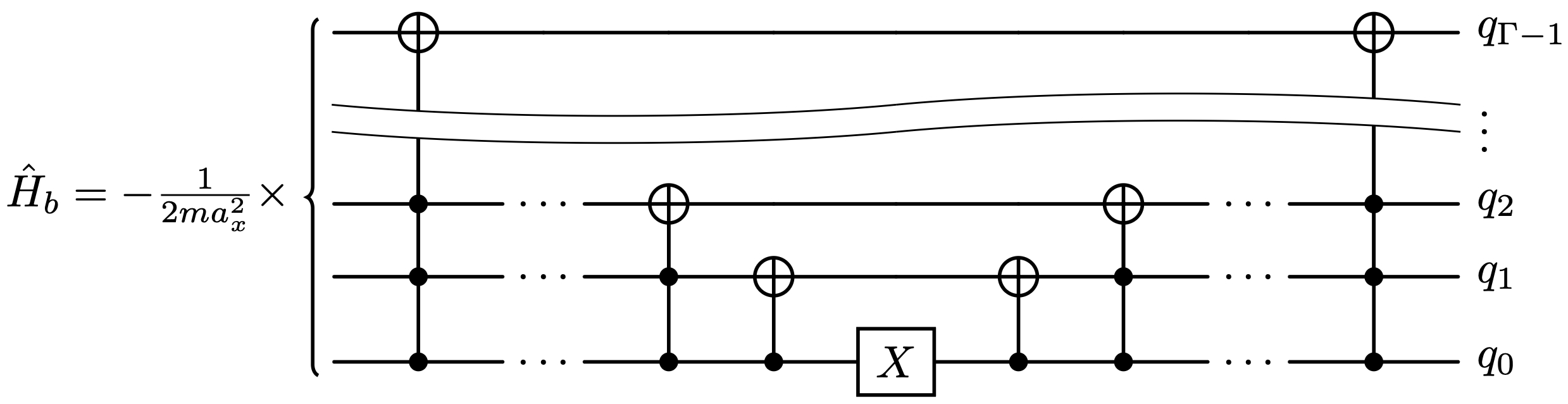}}
\caption{Quantum circuit of $\hat{H}_b$ in Eq.\eqref{HaHbHvcoordinate}.}
\label{FIGHb}
\end{figure}

\paragraph{Hamiltonian matrix in coordinate space: } Let us split the Hamiltonian in Eq.(\ref{Hcoordinate}) into three terms, 
\begin{align}
\hat{H} &= \hat{H}_a +\hat{H}_b + \hat{H}_v, \text{ with} \nonumber \\
\hat{H}_a  & = - \frac{1}{2 m a^2}  \sum_{\alpha = 0}^{\frac{N}{2}-1}     \Bigl (\, | 2 \alpha \rangle \langle 2 \alpha +1 |  + | 2 \alpha +1 \rangle \langle 2 \alpha  |  \,\Bigr ) , \nonumber \\
\hat{H}_b  & = - \frac{1}{2 m a^2}  \sum_{\alpha = 0}^{\frac{N}{2}-1}     \Bigl (\, | 2 \alpha  + 1 \rangle \langle 2 \alpha +2 |  + | 2 \alpha +2 \rangle \langle 2 \alpha  + 1 |  \,\Bigr ) , \nonumber \\
\hat{H}_v  & =  \sum_{\alpha = 0}^{N-1}  \left ( \frac{1}{ m a^2}  + V(x_\alpha ) \right )    | \alpha \rangle \langle \alpha  |  . \label{HaHbHvcoordinate}
\end{align}
The $\hat{H}_a $ and $\hat{H}_b $ terms together are exactly the same as the first two terms in a tight-binding model in Ref.~\cite{Guo:2025xpd}, and their quantum circuits are reproduced here  in Fig.~\ref{FIGHa} and Fig.~\ref{FIGHb} for completeness. The interaction term is given by the sum of all possible $Z$-gate insertions,
\begin{align}
\hat{H}_v & =\frac{1}{ m a^2}  I^{\otimes \Gamma}  \nonumber \\
&  + \frac{V_0}{2 a} \frac{1}{2^{\Gamma-1}} \sum  I_{\Gamma -1}   \otimes (\text{all \ even \ Z  \ insertions})    \nonumber \\
& -  \frac{V_0}{2 a} \frac{1}{2^{\Gamma-1}} \sum Z_{\Gamma -1}  \otimes  (\text{all \ odd \ Z \ insertions})    . \label{HvQC}
\end{align}
 In the second line of Eq.(\ref{HvQC}), even $Z$-gate insertions include zero $Z$ gates.  As an example, with 3 qubits, $\hat{H}_v$ is given by
\begin{align}
\hat{H}_v & =\frac{1}{ m a^2}  I_3 \otimes I_2 \otimes I_1  \nonumber \\
&  + \frac{V_0}{2 a} \frac{1}{4 } ( I_3 \otimes I_2 \otimes I_1  +  I_3 \otimes Z_2 \otimes Z_1   )  \nonumber \\
& - \frac{V_0}{2 a} \frac{1}{4  } (Z_3 \otimes Z_2 \otimes I_1  + Z_3 \otimes I_2 \otimes Z_1 )    .
\end{align}

\begin{figure}[h]
\frame{\includegraphics[width=.59\textwidth]{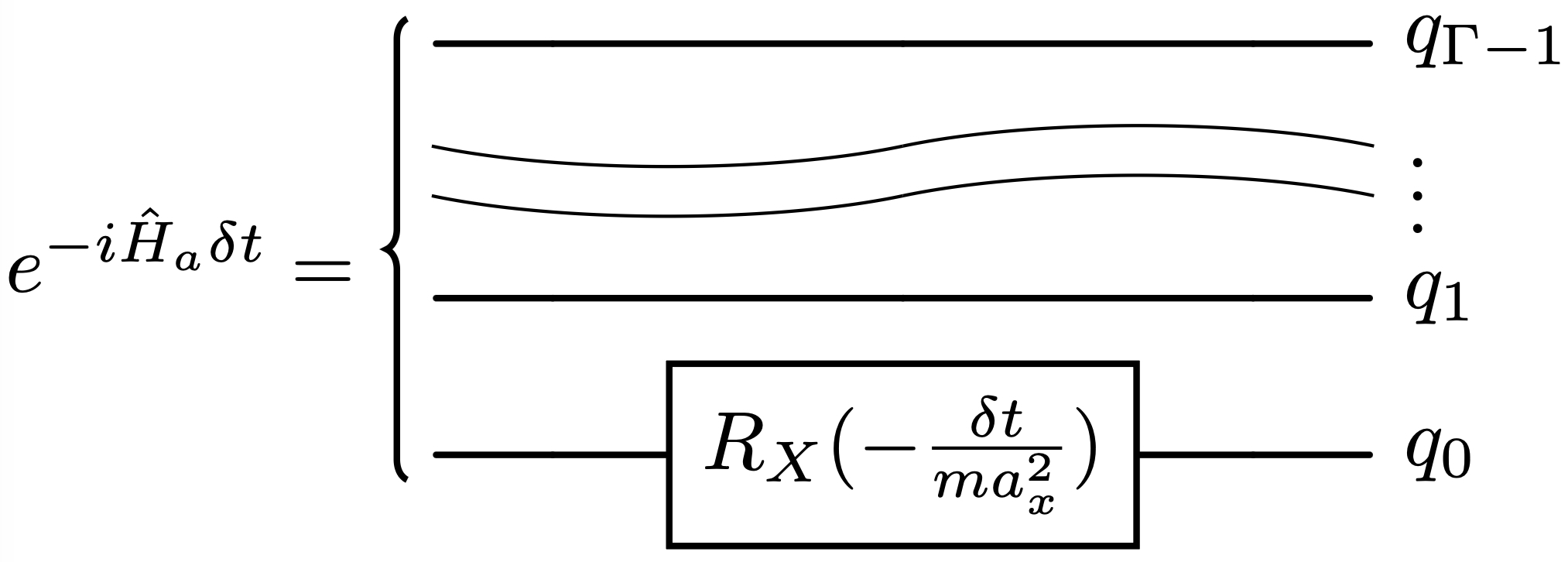}}
\caption{Quantum circuit for time evolution of $e^{- i \hat{H}_a \delta t}$.}
\label{FIGexpHa}
\end{figure}

\begin{figure}[h]
\frame{\includegraphics[width=.99\textwidth]{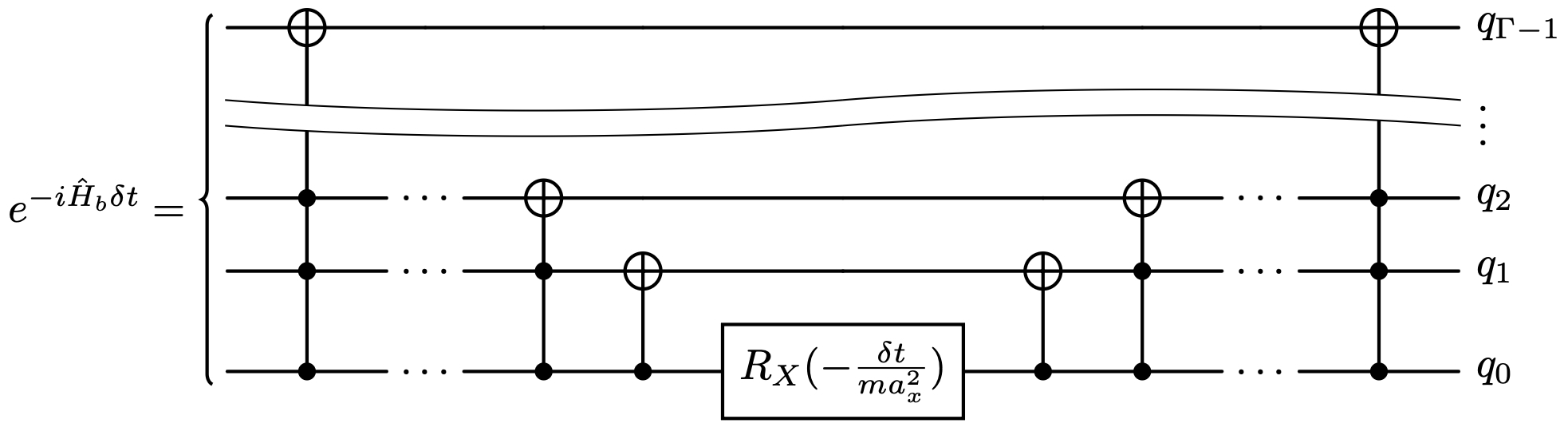}}
\caption{Quantum circuit for time evolution of $e^{- i \hat{H}_b \delta t}$.}
\label{FIGexpHb}
\end{figure}

\begin{figure}[h]
\frame{\includegraphics[width=.59\textwidth]{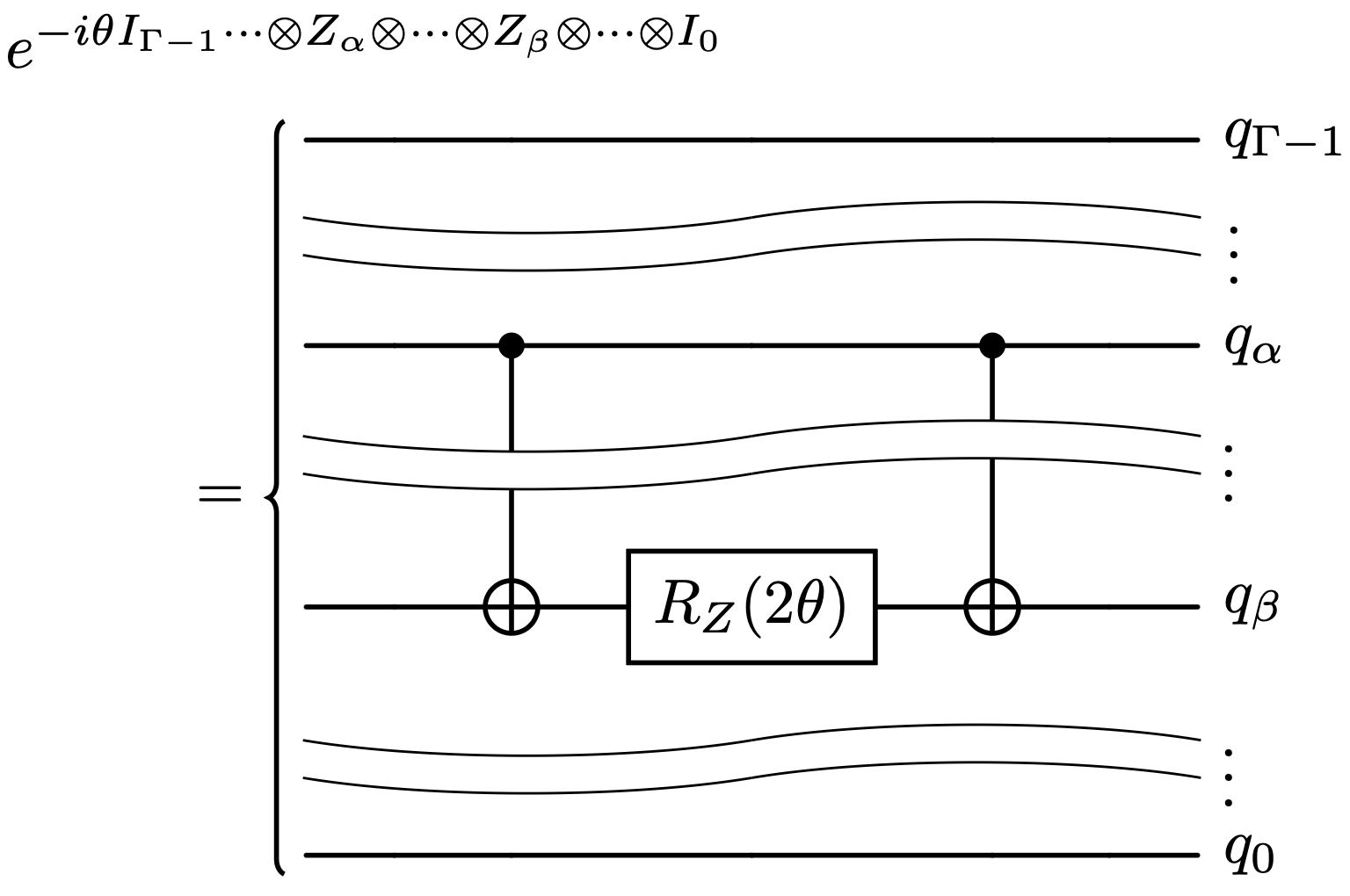}}
\caption{Quantum circuit of a two $Z$-gate insertion term in time evolution of $e^{- i \hat{H}_v \delta t}$:  $e^{-i \theta  I_{\Gamma-1} \cdots \otimes Z_{\alpha} \otimes \cdots   \otimes Z_{\beta} \otimes  \cdots  \otimes I_{0}  } $.}
\label{FIGexpH2zs}
\end{figure}

\begin{figure}[h]
\frame{\includegraphics[width=.89\textwidth]{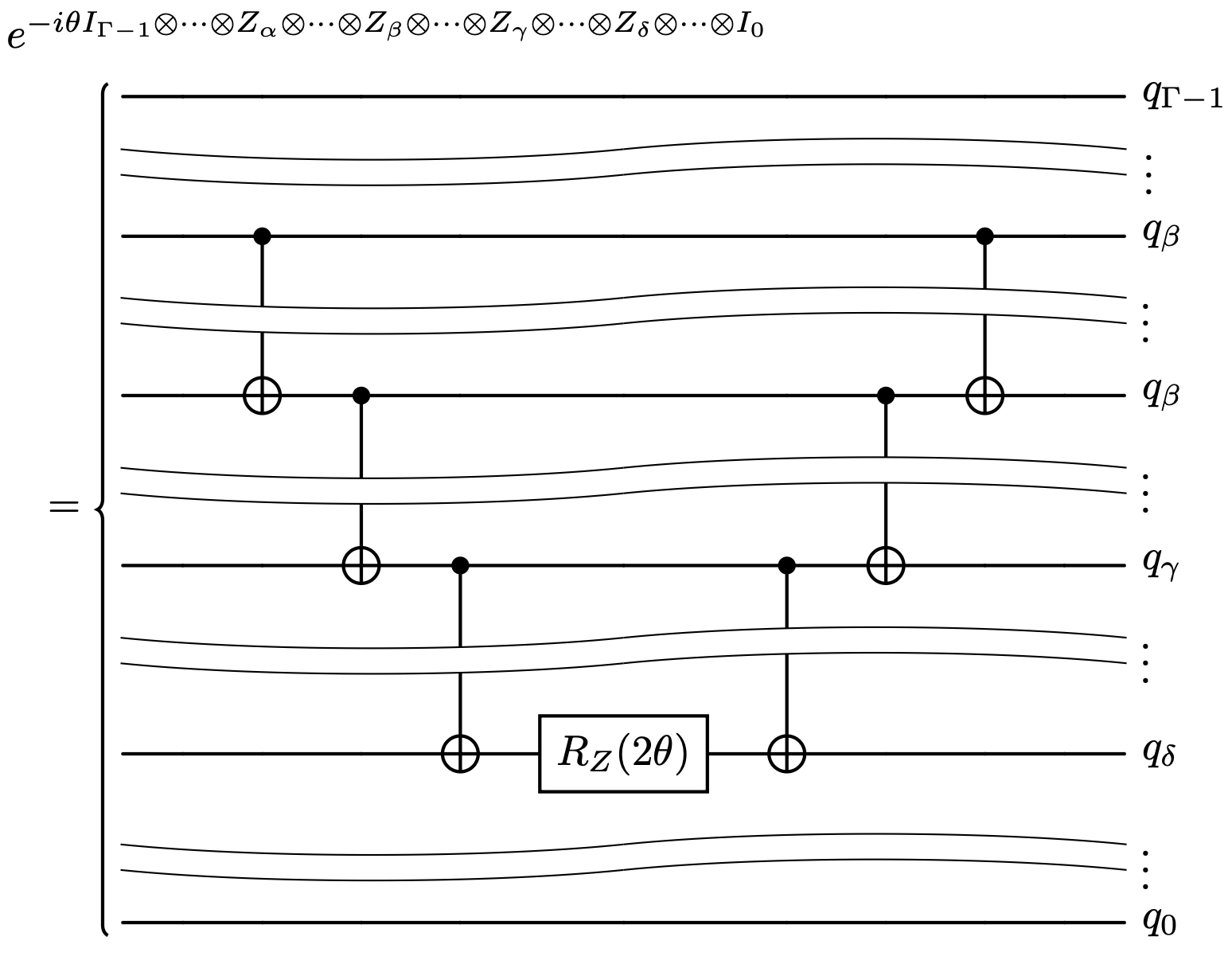}}
\caption{Quantum circuit of  a four $Z$-gate insertion term in time evolution of $e^{- i \hat{H}_v \delta t}$:  $e^{ - i \theta I_{\Gamma-1}  \otimes \cdots \otimes Z_\alpha \otimes \cdots \otimes Z_\beta \otimes \cdots  \otimes Z_\gamma \otimes  \cdots  \otimes Z_\delta \otimes \cdots     \otimes I_0  } $.}
\label{FIGexpH4zs}
\end{figure}

Quantum circuits of time evolutions  $e^{- i \hat{H}_a \delta t}$ and $e^{- i \hat{H}_b \delta t}$ are given in Fig.~\ref{FIGexpHa} and Fig.~\ref{FIGexpHb}, also see details in Ref.~\cite{Guo:2025xpd}. Time evolution of $\hat{H}_v$ is given by,
\begin{align}
e^{- i \hat{H}_v \delta t} &= e^{- i \frac{\delta t}{ m a^2} }  \nonumber \\
& \times  \prod e^{ - i \frac{V_0}{2 a} \frac{\delta t}{2^\Gamma }   I_{\Gamma -1}   \otimes (\text{all even Z  insertions})     } \nonumber \\
& \times \prod e^{  i \frac{V_0}{2 a} \frac{\delta t}{2^\Gamma }   Z_{\Gamma -1}  \otimes  (\text{all odd Z insertions})    } .
\end{align}
Examples for quantum circuits of $e^{- i \hat{H}_v \delta t} $ with two $Z$-gate and four $Z$-gate insertions are shown in  Fig.~\ref{FIGexpH2zs} and Fig.~\ref{FIGexpH4zs}, respectively.

\paragraph{Hamiltonian matrix in momentum basis: }  In the momentum basis, the momentum mode is restricted to a finite size, $k = \frac{2\pi }{L} ( - \frac{N}{2} + n )$, where $n \in [0, N-1] $.   The  Hamiltonian in momentum basis is thus  a finite-sized $N \times N$ matrix.  Splitting the Hamiltonian in Eq.(\ref{Hmomentum}) into two terms, 
\begin{align}
\hat{H} &= \hat{H}_1 + \hat{H}_2, \text{ with } \nonumber \\
\hat{H}_1 & = \sum_{  \alpha  \in  \left [ 0, N-1 \right ] }     \frac{1}{2m } \left ( \frac{2\pi}{L} \right )^2  \left( - \frac{N}{2} +\alpha \right)^2 | \alpha \rangle \langle \alpha  |   , \nonumber \\
\hat{H}_2  & =   \frac{V_0}{L} \sum_{(\alpha, \alpha')  \in  [0, N-1] }        | \alpha' \rangle \langle \alpha |   ,   \label{H1H2momentum}
\end{align} 
where the momentum basis $| k \rangle$ is relabeled to $| \alpha \rangle$ for convenience.
The $N\times N$ diagonal matrix $\hat{H}_1$ and constant matrix $\hat{H}_2$ can be mapped rather straightforwardly to $\Gamma = \log_2 N$ quantum registers. The quantum circuit for $\hat{H}_1$ is given by,
 \begin{equation}
\hat{H}_1 =       \frac{1}{2m } \left ( \frac{2\pi}{L} \right )^2    \hat{U}^2   , 
\end{equation}
where the $\hat{U} $ matrix is defined by
\begin{equation}
 \hat{U}  = \sum_{  \alpha  \in  \left [ 0, N-1 \right ] } \left( - \frac{N}{2} +  \alpha \right) | \alpha \rangle \langle \alpha  |  .
 \end{equation}
 The $\hat{U} $ operator is proportional to the electric field term of the Hamiltonian for a tight-binding model in a constant electric field in  Ref.~\cite{Guo:2025xpd}.  This term  is given in Ref.~\cite{Guo:2025xpd}  by the sum of a constant term and all possible single $Z$-gate insertions,
 \begin{align}
 \hat{U}    = -    \frac{ 1}{2} I^{\otimes \Gamma } -  \sum_{\beta =0}^{\Gamma -1} \frac{2^\beta}{2} I_{\Gamma -1}   \cdots \otimes Z_\beta  \cdots  \otimes   I_0.
 \end{align}
 Hence we find for $\hat{H}_1 $ the quantum circuit, 
  \begin{align}
\hat{H}_1     
 & =    \frac{\left ( \frac{2\pi}{L} \right )^2 }{2m }   \frac{2^{2\Gamma}+4 }{12}  I^{\otimes \Gamma }    \nonumber \\
 &  +  \frac{\left ( \frac{2\pi}{L} \right )^2 }{2m }   \sum_{\alpha =0}^{\Gamma -1} \frac{2^\alpha}{2} I_{\Gamma -1}   \cdots \otimes Z_\alpha  \cdots   \otimes I_0  \nonumber \\
 & + \frac{\left ( \frac{2\pi}{L} \right )^2 }{2m }  \sum_{(\alpha > \beta )=0}^{\Gamma -1}      \frac{ 2^{\alpha+\beta} }{2}      
I_{\Gamma-1}    \cdots \otimes Z_\alpha     \cdots \otimes Z_\beta   \cdots     \otimes I_0   , \label{H1QC}
 \end{align} 
where the last term in Eq.(\ref{H1QC}) is given by the sum of all possible two $Z$-gate insertions. 
 
 The $\hat{H}_2$ is a constant matrix, whose quantum circuit is given by,
\begin{align}
 \hat{H}_2&= \frac{V_0}{L}  I_{\Gamma -1} \otimes \cdots \otimes I_1 \otimes I_0 \nonumber \\
 &+\frac{V_0}{L} I_{\Gamma -1} \otimes \cdots \otimes I_1 \otimes X_0 \nonumber \\
 & + \cdots + \frac{V_0}{L} X_{\Gamma -1} \otimes \cdots \otimes I_1 \otimes I_0 \nonumber \\
 &  + \frac{V_0}{L}  I_{\Gamma -1} \otimes \cdots \otimes X_1 \otimes X_0 \nonumber \\
 & + \cdots  + \frac{V_0}{L}  X_{\Gamma -1} \otimes \cdots \otimes X_1 \otimes X_0  , 
\end{align}
which includes all possible single $X$-gate insertions, two $X$-gate insertions, etc., up to $\Gamma$ $X$-gate insertions.  The matrix elements of $\hat{H}_2$ are given by,
\begin{equation}
\langle \alpha' | e^{ - i \hat{H}_2  \delta t}  | \alpha \rangle  = \begin{cases} 1+  \dfrac{ e^{- i   \frac{ N V_0}{L} \delta t } -1}{N} , & \mbox{if} \ \ \alpha =\alpha' ; \\    \dfrac{ e^{- i  \frac{ N V_0}{L}\delta t } -1}{N}, & \mbox{otherwise} , \end{cases}
\end{equation}
Its time evolution $e^{ - i \hat{H}_V  t} $ is    represented by the quantum circuit,
\begin{align}
e^{ - i \hat{H}_2  \delta  t}  &=\left (1+ \frac{ e^{- i   \frac{N V_0}{L} \delta t } -1}{N}  \right ) I_{\Gamma -1} \otimes \cdots \otimes I_1 \otimes I_0 \nonumber \\
 &+ \frac{ e^{- i   \frac{N V_0}{L} \delta t } -1}{N}  I_{\Gamma -1} \otimes \cdots \otimes I_1 \otimes X_0 \nonumber \\
 & + \cdots + \frac{ e^{- i   \frac{N V_0}{L} \delta t } -1}{N} X_{\Gamma -1} \otimes \cdots \otimes I_1 \otimes I_0 \nonumber \\
 &  + \frac{ e^{- i   \frac{N V_0}{L} \delta t } -1}{N} I_{\Gamma -1} \otimes \cdots \otimes X_1 \otimes X_0 \nonumber \\
 & + \cdots  + \frac{ e^{- i   \frac{N V_0}{L} \delta t } -1}{N}  X_{\Gamma -1} \otimes \cdots \otimes X_1 \otimes X_0  .
\end{align}

\begin{figure}[h]
\frame{\includegraphics[width=.99\textwidth]{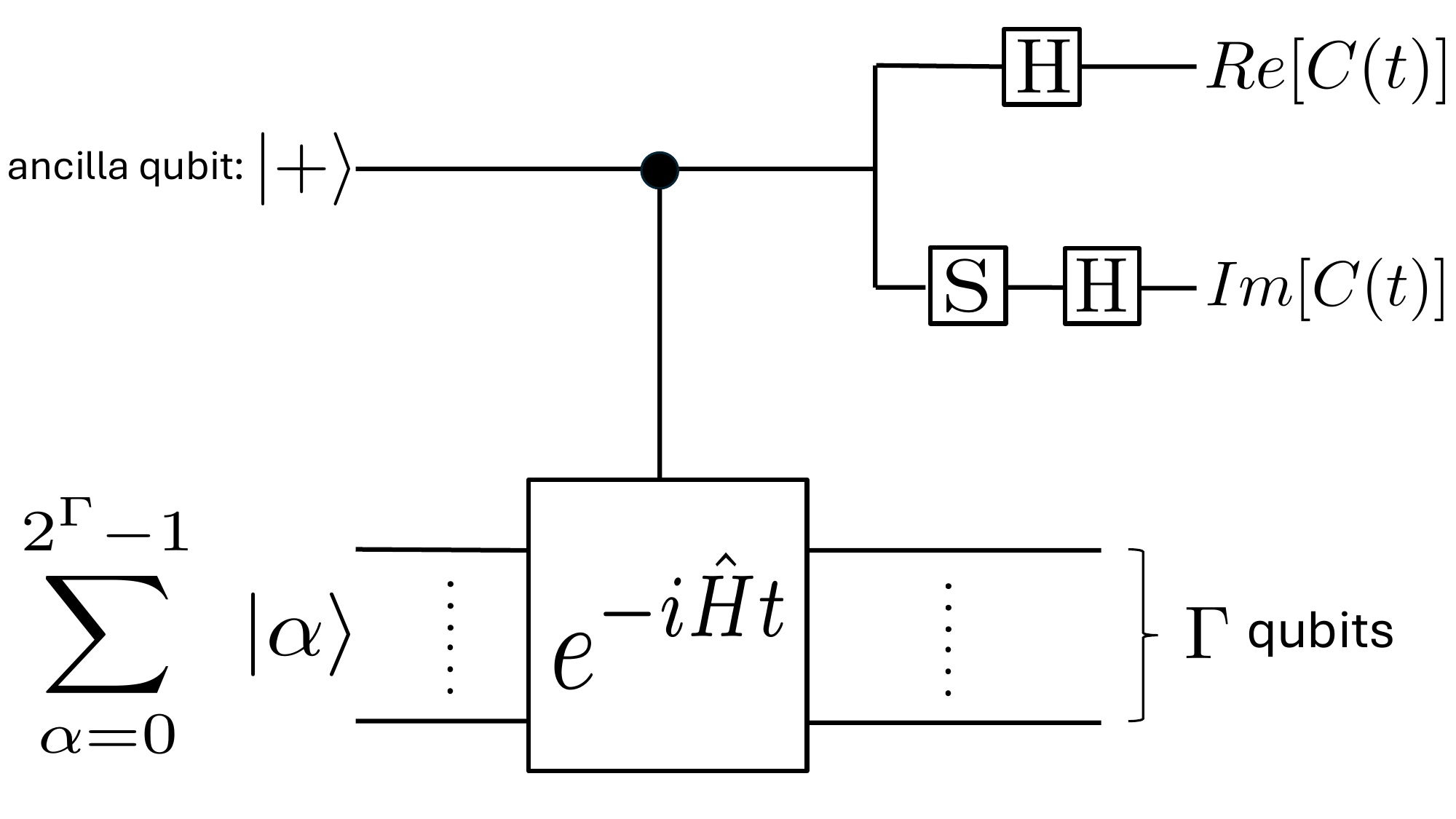}}
\caption{Demo quantum circuit for computing the integrated correlation function $C(t)$ defined in Eq.(\ref{ctdef}). It requires an ancilla qubit to read out the $C(t)$ so the total count of qubits in the implementation is $\Gamma+1$. }
\label{FIGctQC}
\end{figure}

Time evolution of the total Hamiltonian matrix can be  computed via the Trotterization approximation \cite{Trotter1959,Hatano:2005gh}. For example, the lowest order of Trotterization for the Hamiltonian in momentum space is given by,
\begin{equation}
e^{- i \hat{H} \delta t  } \stackrel{ \delta t \rightarrow 0}{\approx }   e^{- i  \hat{H}_{1} \delta t}   e^{- i  \hat{H}_{2} \delta t}     .
\end{equation}
\paragraph{Quantum circuit of integrated correlation functions: } 
Finally, the quantum circuit of computing the ICF,
\begin{equation}
C(t) = Tr \left [e^{ - i \hat{H} t} \right ]  = \sum_{\alpha \in [0, N-1]} \langle \alpha | e^{ - i \hat{H} t } | \alpha \rangle ,\label{ctdef}
\end{equation}
is given in Fig.~\ref{FIGctQC}, also see Fig.4 and Fig.5 in Ref.~\cite{PhysRevD.102.094505}. The Hadamard test method \cite{PhysRevA.75.012328} is used to compute real and imaginary parts of $\langle \alpha | e^{- i \hat{H} t} | \alpha  \rangle$.  The real part of $\langle \alpha | e^{- i \hat{H} t} | \alpha  \rangle$ is given  by $P(0) - P(1)$ in Fig.4 in Ref.~\cite{PhysRevD.102.094505}, where $P(0)$ and $P(1)$ are the probabilities of ancillary qubit at state $|0 \rangle $ and $|1\rangle $ respectively.  The imaginary part of $\langle \alpha | e^{- i \hat{H} t} | \alpha  \rangle$ is given  by $P(1) - P(0)$ in Fig.5 in Ref.~\cite{PhysRevD.102.094505}.

\section{Numerics}\label{sec:numerics}

  \begin{figure}
\includegraphics[width=0.95\textwidth]{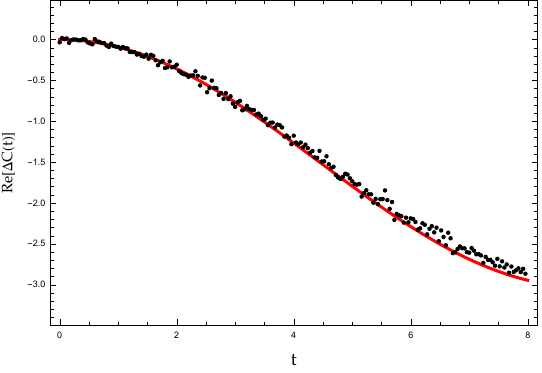}
\includegraphics[width=0.95\textwidth]{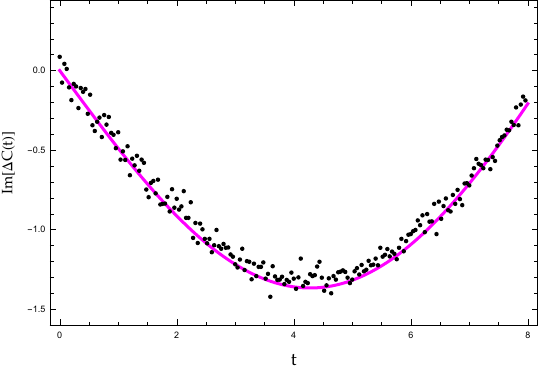}
 \caption{Single-qubit demo plots for real (upper panel) and imaginary (lower panel) parts  of $    \triangle C( t )  = C(t ) - C_0 (t)$  on IBM QPUs (black dots) vs. exact results (solid red and solid purple curves) for a repulsive potential $V_0 =2$ and  other parameters $L=8$, $m=1$, $\delta  t =0.04$. The number of quantum measurements (shots number) is 1000. }  \label{dctonequbitplot} 
 \end{figure}

\paragraph{Single-qubit results:} With a single qubit, the Hamiltonian in coordinate space in Eq.(\ref{Hcoordinate}) is  reduced to the sum of one $X$-gate and one identity gate,
 \begin{equation}
\hat{H}  = - \frac{1}{ m a^2}     X   +  \left( \frac{1}{ m a^2}   +   \frac{V_0}{2 a}  \right)  I   ,
\end{equation}
hence the quantum circuit of time evolution $e^{- i \hat{H} \delta t}$ is given by  product of a $R_X$ gate and a global phase factor,
 \begin{equation}
e^{- i \hat{H} \delta t}= e^{ - i \left( \frac{1}{ m a^2}   +   \frac{V_0}{2 a}  \right) \delta t }     R_X\left( - \frac{2 \delta t}{m a^2} \right) .
\end{equation}
The demo plot of $    \triangle C( t )  = C(t) - C_0 (t)$ on IBM QPUs vs. its exact solutions is shown in Fig.~\ref{dctonequbitplot}.
It involves all terms in the sum in Eq.\eqref{ctdef} for both $C(t)$ and $C_0 (t)$. 
Good overall agreement is observed on a single qubit.

\begin{figure}[h]
\frame{\includegraphics[width=0.99\textwidth]{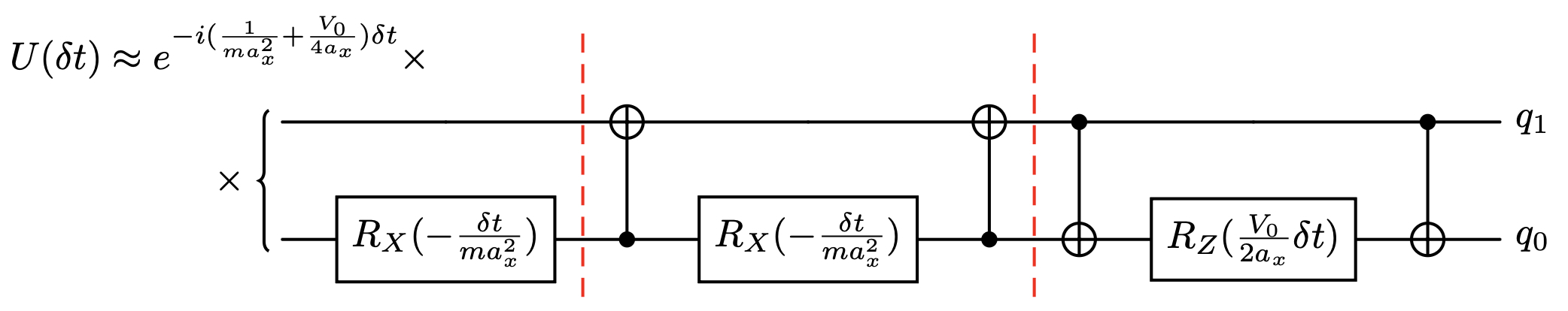}}
\caption{Quantum circuit of the lowest order Trotterization approximation of $U(\delta t) = e^{- i \hat{H} \delta t}$ for two qubits.}
\label{FIG2qubitQC}
\end{figure}

  \begin{figure}
  \centering
\includegraphics[width=0.95\textwidth]{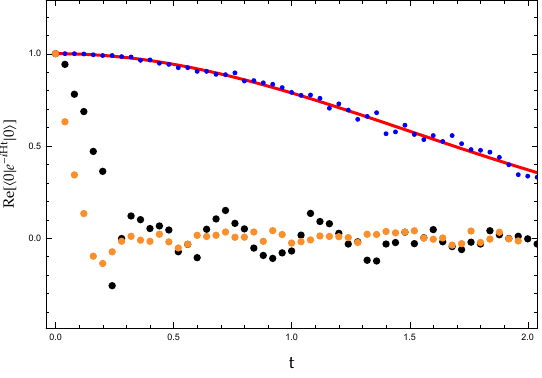}
\includegraphics[width=0.95\textwidth]{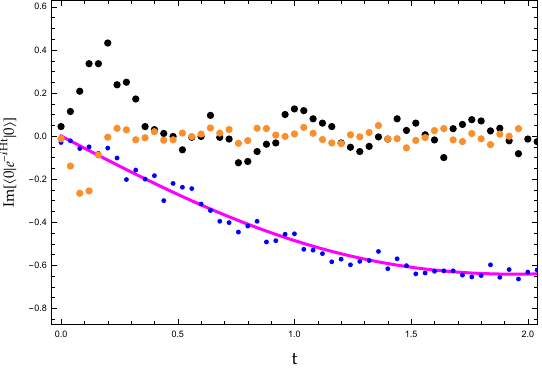}
 \caption{Two-qubit   demo  plots  of real (upper panel) and imaginary (lower panel) parts  of $  \langle \alpha  | e^{ - i \hat{H} t}  | \alpha \rangle  $ with $\alpha=0$  on IBM QPUs  without error mitigation (black dots) and with error mitigation (orange dots) vs. exact results (solid red and solid purple curves) vs. simulator results (blue dots)  for a repulsive potential $V_0 =2$ and parameters $L=5.33$,     $m=1$, $\delta  t =0.04$. The shots number is 1000.  }  \label{twoqubitplot} 
 \end{figure}

\paragraph{Two-qubit results:} For two qubits, the quantum circuit of the lowest order Trotterization approximation of $U(\delta t) = e^{- i \hat{H} \delta t}$ is shown in Fig.~\ref{FIG2qubitQC}. The   demo plots of a single term $  \langle \alpha  | e^{ - i \hat{H} t}  | \alpha \rangle  $ (we picked $\alpha=0$   in Eq.\eqref{ctdef}) on IBM devices  vs. exact results vs. simulated results are shown in  Fig.\ref{twoqubitplot}. We see that the results on Qiskit simulator without errors are consistent with the exact result, but the two-qubit  results (black dots) on IBM devices failed completely, even at the lowest order of Trotterization and for a single term in Eq.\eqref{ctdef}. These measured expectation values collapse rapidly toward zero, losing any of the theoretical behavior, becoming indistinguishable from random noise after only a few Trotter steps. We utilized  the mthree package \cite{PRXQuantum.2.040326} for error mitigation (also see a similar readout error mitigation approach in \cite{PhysRevA.106.012423}), shown as orange dots in Fig.\ref{twoqubitplot}; no  improvement is observed. 

In order to determine the dominant error sources for the Trotterized approach, we ran simulations through the Qiskit Aer noise model simulators separated for each noise channel. Each simulation was performed \(100\) times to build \(95\%\) prediction intervals for where the two-qubit results should land, due to the statistical uncertainty provided by the shots. In the following, we display the simulated noisy computations of  $  Re\left [\langle 0  | e^{ - i \hat{H} t}  | 0 \rangle \right ] $ with three different errors (color-coded bands),  
and compare with the exact result (solid red curve) and the result without error mitigation (black dots).

\begin{figure}[htbp]
    \centering
    \includegraphics[width=0.9\textwidth]{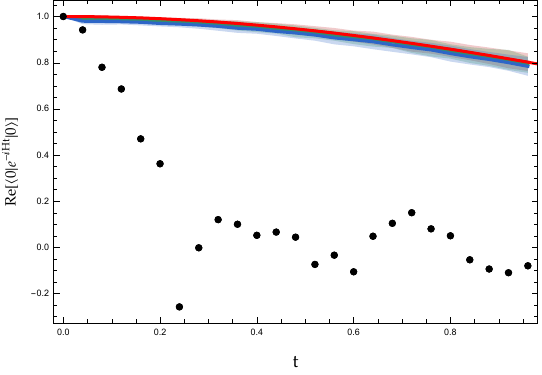}
    \caption{Qiskit Aer simulation of readout errors at \(0.1\%\) (red), \(0.5\%\) (green), and \(1\%\) (blue) in the
    two--qubit result for the real part of \(\langle 0 | e^{-i\hat{H} t} | 0 \rangle\) (top panel in Fig.~\ref{twoqubitplot})
    on IBM QPUs vs.\ exact result (solid red) vs.\ raw measurements without error mitigation (black dots).}
    \label{fig:readout_aer}
\end{figure}

\begin{figure}[htbp]
    \centering
    \includegraphics[width=0.9\textwidth]{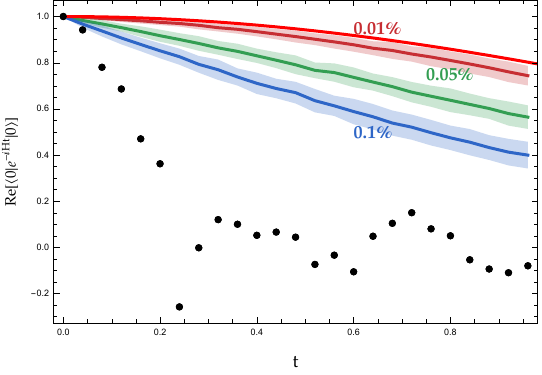}
    \caption{Similar to Fig.~\ref{fig:readout_aer}, but for only single--qubit gate errors at
    \(0.01\%\), \(0.05\%\), and \(0.1\%\).}
    \label{fig:single_aer}
\end{figure}

First we look at the readout error from the ancilla qubit used to output results for $C(t)$, assuming all other qubits operating at perfect fidelity (see Fig.\ref{FIGctQC}). As illustrated in Fig. \ref{fig:readout_aer}, the readout errors are seen to have a negligible impact on the final observable, continuing up to even a \(5\%\) chance of error. This remains consistent with the lack of improvement seen when applying measurement--error mitigation provided by mthree in Fig.~\ref{twoqubitplot}. This behavior is expected, as the readout errors only act on the final measurement stage, and do not affect the unitary evolution itself, whereas this observed empirical decoherence appears to occur during the circuit execution. Similarly, measuring single--qubit gate errors showed only modest deviations from the exact result for the error rates seen on current hardware (see Fig.\ref{fig:single_aer}), without the strong decoherence observed in other simulations.

\begin{figure}[htbp]
    \centering
    \includegraphics[width=0.9\textwidth]{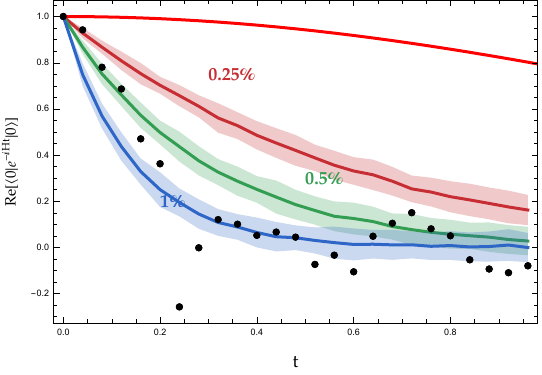}
    \caption{Similar to Fig.~\ref{fig:readout_aer}, but for only two--qubit gate errors at
    \(0.25\%\), \(0.5\%\), and \(1\%\).}
    \label{fig:double_aer}
\end{figure}

In contrast, the two--qubit gate errors begin to show the sharp decoherence patterns that we see when running the demos on hardware. Due to the number of gate operations required by this lowest--order Trotterization, even the most optimistic two--qubit error patterns realistically achievable on current hardware simulate to strongly kill off any coherent dynamics needed for evolution. Because of the multiple entangling gates required by each Trotter step, the total two--qubit gate count grows faster than coherent phase information can be preserved as evolution time increases causing stochastic errors to become substantially more likely. The simulation, Fig. \ref{fig:double_aer}, implies that at least a part of this sharp drop to near random outcomes is coming from this fight between the required relatively long circuit depth compared and the currently achievable coherence times.  Readout errors up to 5\% and single qubit gate errors at typical hardware rates (around 0.01\% -- 0.1\%) produce only minor distortions, while the two--qubit gate errors at realistic values (around 0.25\% -- 1\%) rapidly suppress the coherent evolution once several Trotter steps are applied. These results indicate that, for circuits of the depth considered here, the two--qubit gate errors would need to be reduced to roughly the level of \(0.01\%\) or below for the signal to remain visible within a reasonable deviation at moderate evolution times.  Our noise simulations allow us to estimate the approximate error levels required to preserve a visible signal in the correlator.

\begin{figure}[!htbp]
    \centering
    \includegraphics[width=0.9\textwidth]{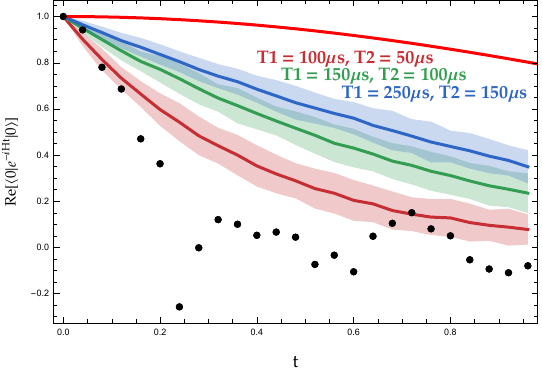}
    \caption{Qiskit Aer simulation of thermal relaxation errors (continued across next panels):
    fixed single--qubit gate length \(50\,\mathrm{ns}\); two--qubit gate length \(100\,\mathrm{ns}\).}
    \label{fig:thermal_aer}
\end{figure}

\begin{figure}[!htbp]\ContinuedFloat
    \centering
    \includegraphics[width=0.9\textwidth]{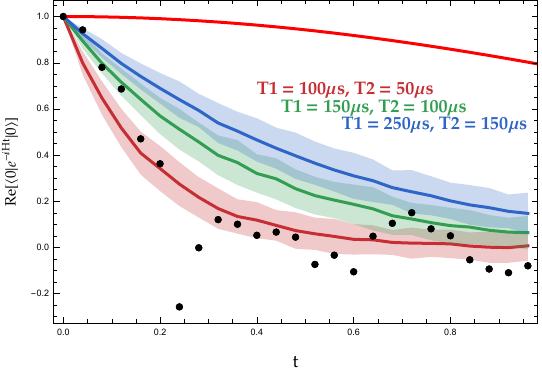}
    \caption{(continued) Two--qubit gate length \(250\,\mathrm{ns}\).}
\end{figure}

\begin{figure}[!htbp]\ContinuedFloat
    \centering
    \includegraphics[width=0.9\textwidth]{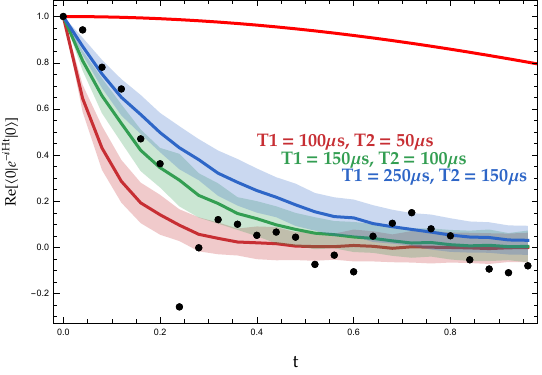}
    \caption{(continued) Two--qubit gate length \(500\,\mathrm{ns}\).
    Thermal parameters swept: \((T_1,T_2)=(100\,\mu s,50\,\mu s)\), \((150\,\mu s,100\,\mu s)\), and
    \((250\,\mu s,150\,\mu s)\).}
\end{figure}

We see this behavior as well when considering thermal relaxation errors in Fig.~\ref{fig:thermal_aer} throughout many different ranges of realistic gate lengths and relaxation times.  The thermal noise simulations similarly indicate that realistic coherence times place strong limits on the usable circuit depth. Even for relatively optimistic parameters \((T1,T2)=(250\mu s,150\mu s)\), the signal decays rapidly once the two--qubit gate durations reach the 100--500ns range typical of current devices, indicating that substantially longer coherence times or shorter entangling gates would be required to preserve coherent evolution over multiple Trotter steps, on the order of 50ns for the two--qubit gates.

\begin{figure}[htbp]
    \centering
    \includegraphics[width=0.9\textwidth]{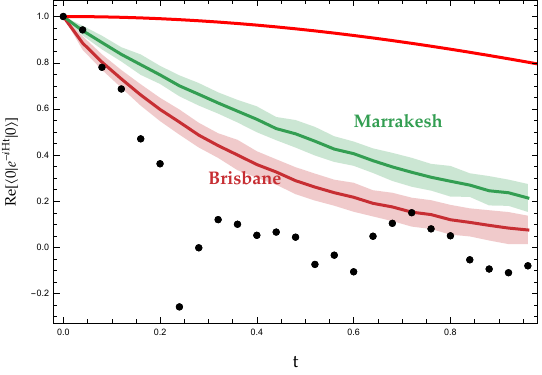}
    \caption{Qiskit Aer simulation of the noise present on a Heron R2 (Marrakesh) and Eagle R3 (Brisbane).}
    \label{fig:backend_aer}
\end{figure}

The choice of noise rates was based on the estimated median noise rates occuring on Heron R2 and Eagle R3 systems, which were observed showing Readout Errors around \(1\%\) and \(2\%\) respectively, Single--Qubit around \(0.02\%\), Two--Qubit around \(0.2\%\) and \(0.8\%\) respectively, and Thermal around \((250\,\mu s, 150\,\mu s)\). IBM also provides combined noise models simulating the actual noise present on their backends. Interestingly, when simulated on a Heron R2 model (IBM Marrakesh) and an Eagle R3 model (IBM Brisbane), the simulated results fall substantially better than the decoherence actually observed in practice, see Fig. \ref{fig:backend_aer}.

These discrepancies between even estimated noise models versus the exact hardware seem to suggest that, while the models capture the average error rates, there may be some underestimated correlated errors or other effects that become significant at the circuit depths required for this type of evolution past the one--qubit setting.

\section{Summary and outlook}\label{sec:summary}

In summary, by using an exactly solvable  1D  quantum mechanical model and the ICF formalism, we studied the feasibility and practicality of extracting scattering phase shift in real time quantum simulation  on current quantum computing architectures. We explored the challenges and limitations facing such a undertaking, including  (1) fast oscillating behavior of integrated correlation functions in real time quantum simulation, and (2) difficulty and failure of quantum simulation encountered on current hardware with even three qubits.

To overcome the fast oscillating behavior of integrated correlation functions in real time quantum simulation, we propose and discuss two mitigation methods: $E+i \varepsilon$ prescription and $ L \rightarrow i L$ rotation in Sec.~\ref{sec:postdataanalysis}. 
Another method employing an averaging technique has been explored previously in Ref.~\cite{Guo:2025vgk}.
The $E+i \varepsilon$ prescription can be applied directly in post data analysis of real-time quantum simulation results, but it requires large volume and large size of Hamiltonian simulation that is clearly not feasible with current hardware. On the other hand, the $ L \rightarrow i L$ rotation method is much more effective  to extract the phase shift directly through the post data analysis, but with the downside that the Hamiltonian matrix becomes non-Hermitian which requires extra efforts to implement on unitary gate operated quantum hardware. Significant progress has been made on implementing imaginary time evolution  and non-Hermitian  Hamiltonian simulation in recent years, see e.g. Refs.~\cite{Motta2020,Nishi2021,PRXQuantum.2.010317,Yi:2025zpa,Mangin-Brinet:2023sey}. How effective these approaches could work with the ICF formalism under current hardware conditions warrants further study.

The numerical tests  carried out on currently available IBM quantum computers are a mixed bag. With two total qubits (an ancilla qubit for measuring $C(t)$ is needed in addition to qubits for real time Hamiltonian evolution, see Fig.\ref{FIGctQC}), the results on IBM hardware are consistent with exact solutions even without any error mitigation. 
However, for three total qubits (one ancilla qubit plus two regular qubits), the results failed miserably with or without error mitigation efforts. Based on the noise simulation models, the Trotterization in its current approach seems poorly suited to current hardware as the circuit length required because time evolution insists upon a substantially higher two--qubit and thermal relaxation fidelity than is currently feasible. Apart from waiting for coherence times to be advanced on the hardware side, some algorithmic considerations, for instance a higher--order Suzuki--Trotter decomposition or Qubitization, may be fruitful in order to reduce the circuit length. There is inherently a fight between the circuit length and depth, as well as the errors arising from noise and coarseness of step size that substantially conspire against fidelity for this implementation.

The formalism and  quantum circuits of 1D quantum mechanical model presented in this work  can be straightforwardly extended to scalar field models. We lay out sufficient details on how it is done in 
Appendix \ref{appendix:scalarmodel}.  
One possible application is a study of few-body scattering in the $\phi^4$ theory  with the ICF formalism on quantum computers.

\acknowledgments
This research is supported by the U.S. National Science Foundation under grant PHY-2418937 (P.G.) and PHY-2531653 (P.G.) and in part Grant No. PHY-2309135 to the Kavli Institute for Theoretical Physics (KITP) (P.G.), and the U.S. Department of Energy under grant  DE-FG02-95ER40907 (F.L.). The work of YZ is supported by the U.S. Department of Energy, Office of Science, Office of Nuclear Physics through Contract No.~DE-AC02-06CH11357.

\appendix

\section{Infinite volume scattering solutions with a contact interaction potential}\label{infvolumedynamics}
The particle scattering  off a contact interaction potential in quantum mechanics can be solved exactly, see e.g. Refs.~\cite{Guo:2016fgl,Guo:2020kph,Guo:2020ikh,Guo:2020spn,Guo:2021uig,Guo:2022row}.  The  1D scattering dynamics in infinite volume is described by Lippmann-Schwinger equation,
\begin{align}
& \psi^{(\infty)}_E (x)   =  e^{ \pm i \sqrt{2 m E} x}   \nonumber \\
& + \int_{-\infty}^{\infty} d x' G^{(\infty)}_0(x-x'; E) V(x')\psi^{(\infty)}_E (x'),  \label{infvolLSeq}
\end{align}
where the free Green's function, $G^{(\infty)}_0(x; E) $, is defined by,
\begin{equation}
G^{(\infty)}_0(x; E)  = \int_{-\infty}^\infty \frac{d p}{2\pi} \frac{e^{i p x}}{E- \frac{p^2}{2m} } = -    \frac{ i  m }{\sqrt{2m E}} e^{i \sqrt{2 m E} |x|} .
\end{equation}
 With a contact potential of $V(x) = V_0 \delta (x)$, Eq.(\ref{infvolLSeq}) is reduced to an algebra equation and can be solved exactly.  
 The analytic solutions of wave function and scattering amplitude  are thus given respectively by,
\begin{equation}
\psi^{(\infty)}_E (x) =  e^{ \pm i \sqrt{2 m E} x} + i f (E) e^{i \sqrt{2 m E} |x|} ,  
\end{equation}
and,
\begin{equation}
f(E) = - \frac{ m V_0 }{\sqrt{2 m E} + i m V_0} .
\end{equation}
The scattering amplitude is typically parameterized by the scattering phase shift,
\begin{equation}
f(E)  = \frac{e^{2 i \delta(E)} -1}{2i} = \frac{1}{\cot \delta (E) - i},
\end{equation}
which leads to the solution of scattering phase shift for a contact interaction potential,
\begin{equation}
\delta(E) = \cot^{-1} \left ( - \frac{\sqrt{2 m E} }{m V_0} \right ).
\end{equation}  

The transmission amplitude $T(E)$ is defined through the forward scattering solution by,
\begin{equation}
T(E) = 1+ i f(E)  = \cos \delta(E) e^{i \delta (E)}= \frac{ \sqrt{2 m E} }{\sqrt{2 m E} + i m V_0} .
\end{equation}
The Muskhelishvili-Omn\`es  dispersion integral representation of $T(E)$ is given in Appendix \ref{Fridelformula}.

\section{Friedel formula and Krein's theorem in infinite volume potential scattering theory}\label{Fridelformula} 
In this appendix,  we briefly  summarized some key results of   Friedel formula  and Krein's theorem in infinite volume scattering theory,  see more details in  Refs.~\cite{PhysRev.187.345,Guo:2022row}.

 First of all, considering that a non-relativistic spinless particle of mass $m$ is scattered off a  potential barrier, the dynamics is   described by Schr\"odinger equation,
 \begin{equation}
 \hat{H}^{(\infty)} | \Psi^{(\infty)}_E \rangle  = E  | \Psi^{(\infty)}_E   \rangle  , \ \ \ \  \hat{H}^{(\infty)}=\hat{H}^{(\infty)}_0 + \hat{V}, \label{schrodingereq}
 \end{equation}
 where $\hat{H}^{(\infty)}_0 =  - \frac{1}{2 m} \frac{d^2}{d x^2}$  and $\hat{V}$  stand for free Hamiltonian  and  potential operators of the   system respectively.

\paragraph{$S$-matrix and Moller operators in  potential scattering theory: }  
 The $S$-matrix operators  is defined through Moller operators by
 \begin{equation}
 \hat{S}(E) = \hat{\Omega}^{\dag}_{E - i 0} \hat{\Omega}_{E + i 0} ,   \label{SMollerdef}
 \end{equation}
 where Moller operator  evolve the  wave function of non-interacting system, $| \Psi^{(\infty, 0)}_E \rangle$,   into the wave function of a interacting system, $| \Psi^{(\infty)}_E \rangle$, by
 \begin{equation}
 | \Psi^{(\infty)}_E \rangle =\hat{ \Omega}_{E }  | \Psi^{(\infty, 0)}_E \rangle .
 \end{equation}
Moller operators, $\hat{ \Omega}_{E+i0 }$ and $\hat{\Omega}_{E-i0 }$,    describe systems evolve   forward in time and   backward in time    respectively, and conservation of probability yields
\begin{equation}
 \hat{\Omega}^\dag_{E } \hat{\Omega}_{E }     = 1 ,  \ \ \ \ \hat{S}^\dag (E)     \hat{S}(E)  =  1.  \label{normMollerops}
\end{equation}

\paragraph{Formal scattering theory: }  The Lippmann-Schwinger  (LS) equation for a scattering system
\begin{equation}
 | \Psi^{(\infty)}_E \rangle  =  | \Psi^{(\infty, 0)}_E \rangle +  \hat{G}^{(\infty)} (E) \hat{V} | \Psi^{(\infty, 0)}_E \rangle  
\end{equation}
yields
\begin{equation}
 \hat{ \Omega}_{E }  = 1+   \hat{G}^{(\infty)} (E) \hat{V},   \label{OmegaGV}
\end{equation}
where  $ \hat{G}^{(\infty)} (E) $ stands for   full Green's function operator that satisfies Dyson equation,
\begin{equation}
\hat{G}^{(\infty)} (E) = \hat{G}^{(\infty)}_0 (E) + \hat{G}^{(\infty)}_0 (E) \hat{V}\hat{G}^{(\infty)} (E) = \frac{1}{E- \hat{H}^{(\infty)}} . \label{Dysoneq}
\end{equation}
The  $ \hat{G}^{(\infty)}_0 (E)  = \frac{1}{E- \hat{H}^{(\infty)}_0 }$ stands for the free Green's function  for non-interacting system. The normalization relation of Moller operator also yields a useful relation
\begin{equation}
 \hat{\Omega}^\dag_{E }    =   \hat{\Omega}^{-1}_{E }  =  1 -   \hat{G}^{(\infty)}_0 (E) \hat{V}. \label{OmegaInverse}
\end{equation}

\paragraph{Friedel formula and Krein's theorem  in potential scattering theory: }   Using all the relations listed above in this section, with some manipulations, see details in  \cite{PhysRev.187.345,Guo:2022row}, we find
 \begin{equation}
  Tr \left [     \hat{ \Omega}_{E  }   \frac{d}{d E} \hat{\widetilde{\Omega}}^{\dag}_{E  }  - \hat{\widetilde{\Omega}}^{\dag}_{E  }   \frac{d}{d E}  \hat{ \Omega}_{E  }    \right ]     = 2 \; Tr \left [ \hat{G}^{(\infty)} (E) -  \hat{G}^{(\infty)}_0 (E)     \right ] , \label{OmegadiffG}
\end{equation}
where the trace  is defined by the sum or integral of specific basis, e.g.  in coordinate space basis, the trace of Green's function operator  is defined by
\begin{equation}
   Tr \left [ \hat{G}^{(\infty)} (E)     \right ]  = \int_{-\infty}^\infty  d x  \langle x|  G^{(\infty)} ( E)   | x \rangle .
\end{equation}
The Eq.(\ref{SMollerdef}) and Eq.(\ref{OmegadiffG})  thus yield  the  Friedel  formula for   a scattering system: 
 \begin{align}
& \frac{1}{4 i}  Tr \left [ \hat{S}^\dag (E)   \frac{d}{d E}   \hat{S}(E)  -  \hat{S}(E)  \frac{d}{d E}   \hat{S}^\dag (E)   \right ] \nonumber \\
& = - Im \left ( Tr     \left [  \hat{G}^{(\infty)} (E) -  \hat{G}^{(\infty)}_0 (E)     \right ]  \right ) . \label{Friedelcomplexpot}
\end{align}

Using unitarity relation of $S$-matrix   and identity $$Tr \left [ \ln \hat{S}(E ) \right ] = \ln \det \left [ \hat{S}(E) \right ],$$
we can rewrite the Friedel formula  in a more compact form
 \begin{align}
 &  \frac{1}{2 i}         \frac{d}{d E}  \ln \left ( \det \left [   \hat{S}(E)   \right ]  \right )    = \frac{d}{d E} Tr[ \delta(E)]  \nonumber \\
 &  =- Im \left ( Tr     \left [  \hat{G}^{(\infty)} (E) -  \hat{G}^{(\infty)}_0 (E)     \right ]  \right )  ,  \label{DiscGlndetScompact}
\end{align}
where $\delta(E)$ refers to the diagonal matrix of scattering amplitudes, which is related to $S$-matrix by $\hat{S}(E) = e^{2 i \delta(E)}$.

Assuming both Green's function and $S$-matrix having the branch cuts along real   axis in complex $E$-plane,   the Green's function can be constructed by Cauchy's integral through the imaginary part  of Green's function across branch  cuts. 
\begin{equation}
 \hat{G}^{(\infty)}(E)  = \frac{1}{\pi}  \int_0^\infty   d \epsilon \frac{ Im     \hat{G}^{(\infty)}(\epsilon)  }{\epsilon - E}. \label{dispG}
\end{equation}
The Eq.(\ref{dispG}) and Eq.(\ref{DiscGlndetScompact}) together yield 
  \begin{equation}
  Tr \left [    \frac{1}{E- \hat{H}^{(\infty)} }  -   \frac{1}{E- \hat{H}^{(\infty)}_0 } \right ] =  -    \frac{1}{\pi}  \int_0^\infty   d \epsilon  \frac{  Tr \left [ \delta (  \epsilon)  \right ]  }{(\epsilon- E)^2}  . \label{Kreintheorem}
\end{equation}
 The relations in Eq.(\ref{DiscGlndetScompact}) and  Eq.(\ref{Kreintheorem}) are also referred as Friedel formula \cite{doi:10.1080/00018735400101233,Friedel1958}  in condensed matter physics and  Krein's theory  \cite{krein1953trace,BirmanKrein1962} in spectral theory respectively.

\paragraph{Muskhelishvili-Omn\`es (MO) representation of Krein's theorem: }  The $S$-matrix can also be related to Fredholm determinant, see e.g. Refs.~\cite{GOL64,PhysRev.103.1565},
\begin{equation}
D(E) = \det \left [ 1 - \hat{G}^{(\infty)}_0 (E) \hat{V} \right ]
\end{equation}
by
\begin{equation}
\det \left [ \hat{S}(E)  \right ]= e^{2 i  Tr \left [ \delta (E) \right ] } = \frac{D (E- i 0)}{ D(E+i 0)} .
\end{equation}
The solution of Fredholm determinant is given by Muskhelishvili-Omn\`es  dispersion integral representation  \cite{Omnes:1958hv,PhysRevD.16.896} by
\begin{equation}
D(E) =  e^{- \frac{1}{\pi} \int_0^\infty d \epsilon \frac{Tr\left [ \delta (\epsilon) \right ] }{\epsilon-E}} .
\end{equation} 
In 1D, the transmission amplitude $T (E)$ can be  identified as inverse of Fredholm determinant, see e.g. Refs.~\cite{Guo:2022row,Guo:2022jyk,Guo:2024bar},
\begin{equation}
 T(E)   = \frac{1}{D(E) } =  e^{   \frac{1}{\pi} \int_0^\infty d \epsilon \frac{Tr\left [ \delta (\epsilon) \right ] }{\epsilon-E}}  .
\end{equation}
Using relation in Eq.(\ref{Kreintheorem}), we also find another representation of Krein's theorem:
   \begin{equation}
-    \frac{d}{d E} \ln  \left [ T(E) \right ]   =  Tr \left [    \frac{1}{E- \hat{H}^{(\infty)} }  -   \frac{1}{E- \hat{H}^{(\infty)}_0 } \right ]   . \label{MOrepKreintheorem}
\end{equation}

\section{Extension to scalar field theory models}\label{appendix:scalarmodel}

The quantum circuits presented in Sec.~\ref{sec:QC} can be readily applied to the scalar field theory models, such as $\phi^4$ theory.  The discretized Hamiltonian of a real scalar field $\phi$ in one spatial dimension is given e.g. in Ref.~\cite{PhysRevA.99.052335,qr72-51v1} by
\begin{align}
\hat{H} & = a \sum_{j \in [ 0, N_x -1 ]} \bigg [ \frac{1}{2} \hat{\Pi}^2 (x_j) +  \frac{1 }{2} m^2  \hat{\phi}^2 (x_j)   \nonumber \\
& +   \frac{1}{2 a^2}  \left ( \hat{\phi}(x_{j+1}) - \hat{\phi}(x_j )  \right )^2  + \frac{\lambda }{4!}   \hat{\phi}^4(x_j)   \bigg ],
\end{align}
where the periodic boundary condition is assumed
\begin{equation}
\hat{\phi} (x_0) = \hat{\phi} (x_{N_x} ) .
\end{equation}
The    discretized  positions   are defined by  $x_j =  a j $, where $ j \in [0, N_x-1]$. The  $a = \frac{L}{N_{x}}$ is lattice spacing, and $L$ is the size of periodic box along spatial extent. The $m$ and $\lambda$ refer to the mass and coupling strengths of scalar field  $\hat{\phi}$ respectively. The conjugate momenta  operator $\hat{\Pi} (x_j )$  is given by 
\begin{equation}
\hat{\Pi} (x_j ) = - i \frac{\partial}{\partial \phi (x_j) } .
\end{equation}
Next, we also discretize the value of $\phi (x_j) $ field at $j$-th site to
\begin{equation}
\phi_{\alpha_j}  = - \frac{\phi_{max}}{2} + a_{\phi} \alpha_j, \ \ \  \alpha_j \in [0, N_\phi  -1], \label{phijdef}
\end{equation} 
where $a_\phi = \frac{\phi_{max}}{N_\phi -1}$ is spacing of discretized $\phi$ field. The subscript $j$ is added in $\alpha_j$ to label the  value of $\phi_\alpha$ at $j$-th site $x_j$.

\paragraph{Construction of Hamiltonian matrix in discretized $\phi$ field basis: } On $j$-th site, the basis that is used to construct the Hamiltonian matrix can be chosen as discrete $\phi$ field basis
\begin{equation}
\hat{\phi} (x_j) | \alpha_j \rangle  = \phi_{\alpha_j}  | \alpha_j \rangle ,
\end{equation}
hence
\begin{equation}
\hat{\Pi} (x_j )  | \alpha_j \rangle  = - i \frac{ | \alpha_j +1 \rangle - | \alpha_j \rangle  }{a_\phi} ,
\end{equation}
and
\begin{equation}
\hat{\Pi}^2 (x_j )  | \alpha_j \rangle  = -  \frac{ | \alpha_j +1 \rangle  - 2  | \alpha_j \rangle  +| \alpha_j -1 \rangle  }{a^2_\phi} .
\end{equation}
 The   basis of full Hamiltonian can be constructed by tensor product of discrete $\phi$ field basis at each site,
\begin{equation}
 | \alpha_{N_\phi -1} , \cdots,  \alpha_1 ,   \alpha_0 \rangle   =  | \alpha_{N_\phi -1} \rangle \otimes \cdots  | \alpha_1 \rangle \otimes  | \alpha_0 \rangle .
\end{equation}
The Hamiltonian matrix can be split into sum of local term at each site and interaction between nearest neighbor sites,
\begin{align}
\hat{H} & =a \sum_{j \in [0, N_x-1]} I_{N_{\phi}-1}  \cdots \otimes \hat{H}^{(local)}_{j}  \cdots  \otimes I_0 \nonumber \\
& + \frac{a}{a_\phi^2} \sum_{j \in [0, N_x-1]} I_{N_{\phi}-1}  \cdots \otimes \hat{\phi} (x_{j+1} )  \otimes  \hat{\phi} (x_{j} ) \cdots  \otimes I_0 , \label{Htotphidef}
\end{align}
where
\begin{equation}
\hat{H}^{(local)}_{j}   =  \frac{1}{2} \hat{\Pi}^2 (x_j) +  \frac{1 }{2}  \left ( m^2 + \frac{2}{a_\phi^2} \right )  \hat{\phi}^2 (x_j)  +  \frac{\lambda }{4!}   \hat{\phi}^4(x_j)  .
\end{equation}
The matrix of $ \hat{\phi} (x_{j} ) $ at $j$-th site in terms of $\phi$ field basis is given by
\begin{equation}
\hat{\phi} (x_{j} )  = \sum_{\alpha_j \in [0, N_\phi -1]} \phi_{\alpha_j} | \alpha_j \rangle \langle \alpha_j | , \label{phijmatrix}
\end{equation}
where  discrete $\phi_{\alpha_j}$ values are defined in Eq.(\ref{phijdef}).
The $\hat{H}^{(local)}_{j}  $ matrix is given by
\begin{equation}
\hat{H}^{(local)}_{j}   = \hat{H}_j^{(a+b)}  + \hat{H}_j^{(ho)}  + \hat{H}_j^{(v)} ,
\end{equation}
where
\begin{align}
\hat{H}_j^{(a+b)} & = - \frac{1}{2m a_\phi^2} \sum_{\alpha_j = 0}^{N_\phi -1} \Bigl (\, | \alpha_j \rangle \langle \alpha_j +1 | + | \alpha_j +1 \rangle \langle \alpha_j  | \,\Bigr ), \nonumber \\
\hat{H}_j^{(ho)} & = \sum_{\alpha_j = 0}^{N_\phi -1}  \left [ \frac{1}{m a_\phi^2} + \frac{1}{2} (m^2 +  \frac{2}{a_\phi^2} ) \phi_{\alpha_j}^2 \right ]  | \alpha_j \rangle \langle \alpha_j  |, \nonumber \\
\hat{H}_j^{(v)} & = \frac{\lambda }{4!}   \sum_{\alpha_j = 0}^{N_\phi -1}    \phi_{\alpha_j}^4    | \alpha_j \rangle \langle \alpha_j  | . \label{Hphijdef}
\end{align}
This Hamiltonian has the same structure as the one in Eq.\eqref{Hcoordinate} for quantum mechanics so all the infrastructure developed in quantum mechanics  can be straightforwardly transcribed over to the $\phi^4$ theory.

\paragraph{Quantum circuits of $\phi$ field Hamiltonian: } The $\phi$ field Hamiltonian in Eq.(\ref{Htotphidef}) can be mapped onto  $N_x$ sets of quantum registers  with $\Gamma_\phi = \log_2 N_\phi$ qubits for each site, the total number of quantum registers are $N_x \times \Gamma_\phi $.

The $\hat{\phi} (x_j)$ matrix in Eq.(\ref{phijmatrix}) at $j$-th site resembles the Electric field term in the tight-binding model under a uniform electric field, see e.g. Ref.~\cite{Guo:2025xpd}. The quantum circuit of $\hat{\phi} (x_j)$ matrix is given in Ref.~\cite{Guo:2025xpd} and also in Sec.~\ref{sec:QC},
\begin{equation}
\hat{\phi} (x_{j} )  =  a_\phi  \sum_{\alpha_j \in [0, N_\phi -1]}  \left( - \frac{N_\phi -1}{2} + \alpha_j \right) | \alpha_j \rangle \langle \alpha_j | = a_\phi \hat{U}_\phi ,  
\end{equation}
where 
\begin{equation}
\hat{U}_\phi   =    -  \sum_{\beta =0}^{\Gamma -1} \frac{2^\beta}{2} I_{\Gamma -1}   \cdots \otimes Z_\beta  \cdots  \otimes   I_0 . \label{Uphidef}
\end{equation}
The general few-particle creation and annihilation operators  hence can be constructed via the tensor product of $\hat{U}_\phi $ operators, for instance, a two-particle creation operator that create two particles at $x_1$ and $x_2$ respectively  can be constructed by 
\begin{equation}
\mathcal{O}^\dag (x_1, x_2)   =  I_{N_\phi -1}  \cdots \otimes  \hat{ \phi } (x_1) \otimes  I_{\alpha_1 -1}  \cdots    \otimes  \hat{ \phi } (x_2)    \cdots \otimes I_0.
\end{equation}

The quantum circuit of $\hat{H}_j^{(a+b)}$ is given in Fig.~\ref{FIGHa} and Fig.~\ref{FIGHb}. The quantum circuits of $\hat{H}_j^{(ho)} $ and $\hat{H}_j^{(v)}$ can be constructed through quantum circuit of $\hat{U}_\phi  $ in Eq.(\ref{Uphidef}) by
\begin{align}
\hat{H}_j^{(ho)} & =  \frac{1}{m a_\phi^2}  I^{\otimes \Gamma_\phi } +\left( 1+  \frac{1}{2}  a^2_\phi m^2   \right)  \hat{U}^2_\phi  , \nonumber \\
\hat{H}_j^{(v)} & = \frac{a^4_\phi  \lambda }{4!}    \hat{U}^4_\phi    .  
\end{align}

\bibliography{ALL-REF.bib}

\end{document}